\documentclass[fleqn,usenatbib,useAMS]{mnras}

\usepackage{graphicx}	
\usepackage{amsmath}	
\usepackage{amssymb}	
\usepackage{multicol}        
\usepackage{xcolor}

\usepackage{bm}		
\usepackage{pdflscape}	
\usepackage{siunitx}

\newcommand{\unitvec}{\ensuremath{\hat{\bm{n}}}}
\newcommand{\dx}[1]{\mathrm{d}{#1}\,}
\newcommand{\dxsq}[1]{\mathrm{d}^2{#1}\,}
\newcommand{\Cltot}[1]{C_{#1}^{\mathrm{tot}}}
\newcommand{\vecl}{\bm{l}}

\newcommand{\vecL}{\bm{L}}

\newcommand{\Aksz}{A_{\mathrm{A16}}}

\newcommand{\zmid}{z_{\mathrm{mid}}}

\newcommand{\lmin}{l_{\mathrm{min}}}

\newcommand{\lmax}{l_{\mathrm{max}}}

\newcommand{\planck}{{\it Planck}}

\newcommand{\Tcib}{T_{\mathrm{CIB}}}

\newcommand{\websky}{\textsc{websky}}
\newcommand{\sehgal}{S10}
\newcommand{\nemo}{\textsc{nemo}}

\newcommand\eqn[1]{equation~\ref{#1}}

\newcommand\fig[1]{Figure~\ref{#1}}

\newcommand\sect[1]{Section~\ref{#1}}

\newcommand\app[1]{Appendix~\ref{#1}}

\newcommand\NM[1]{\textcolor{red}{NM: #1}}

\DeclareMathSizes{12}{17.28}{9}{7} 

\usepackage[T1]{fontenc}
\usepackage{ae,aecompl}

\usepackage{newtxtext,newtxmath}

\title[Limits on the kSZ trispectrum from ACT DR6]{The Atacama Cosmology Telescope: Reionization kSZ trispectrum methodology and limits}

\author[N~MacCrann et al.]{Niall~MacCrann$^{1,2}$\thanks{Contact email: \href{nm746@cam.ac.uk}{nm746@cam.ac.uk}},
Frank~J.~Qu$^{1,2}$,
Toshiya~Namikawa$^{3}$,
Boris~Bolliet$^{4,5}$,
Hongbo~Cai$^{6}$,
\newauthor
Erminia~Calabrese$^{7}$,
Steve~K.~Choi$^{8}$,
Omar~Darwish$^{9}$,
Simone~Ferraro$^{10}$,
Yilun~Guan$^{11}$,
J.~Colin Hill$^{12}$,
\newauthor
Matt~Hilton$^{13,14}$,
Ren\'ee~Hlo\v{z}ek$^{15,16}$,
Darby~Kramer$^{17}$,
Mathew~S.~Madhavacheril$^{18}$,
\newauthor
Kavilan~Moodley$^{19,20}$,
Neelima~Sehgal$^{21}$,
Blake~D.~Sherwin$^{1,2}$,
Crist\'obal Sif\'on$^{22}$,
Suzanne~T.~Staggs$^{23}$,
\newauthor
Hy~Trac$^{24}$,
Alexander~Van~Engelen$^{17}$,
Eve~M.~Vavagiakis$^{25}$,
\\
$^{1}$DAMTP, Centre for Mathematical Sciences, University of Cambridge, Wilberforce Road, Cambridge CB3 OWA, UK\\
$^{2}$Kavli Institute for Cosmology Cambridge, Madingley Road, Cambridge CB3 0HA, UK\\
$^{3}$Center for Data-Driven Discovery, Kavli IPMU (WPI), UTIAS, The University of Tokyo, Kashiwa, 277-8583, Japan\\
$^{4}$Astrophysics Group, Cavendish Laboratory, J. J. Thomson Avenue, Cambridge CB3 0HE, United Kingdom\\
$^{5}$Kavli Institute for Cosmology, University of Cambridge, Madingley Road, Cambridge CB3 0HA\\
$^{6}$Department of Physics and Astronomy, University of Pittsburgh, Pittsburgh, PA, USA 15260\\
$^{7}$School of Physics and Astronomy, Cardiff University, The Parade, Cardiff, Wales CF24 3AA, UK\\
$^{8}$Department of Physics and Astronomy, University of California, Riverside, CA 92521, USA\\
$^{9}$Universit\'e de Gen\`eve, D\'epartement de Physique Th\'eorique et CAP, 24 Quai Ansermet, CH-1211 Gen\`eve 4, Switzerland\\
$^{10}$Lawrence Berkeley National Laboratory, One Cyclotron Road, Berkeley, CA 94720, USA\\
$^{11}$Dunlap Institute for Astronomy and Astrophysics, University of Toronto, Toronto, ON M5S 3H4, Canada\\
$^{12}$Department of Physics, Columbia University, New York, NY, USA 10027\\
$^{13}$Wits Centre for Astrophysics, School of Physics, University of the Witwatersrand, Private Bag 3, 2050, Johannesburg, South Africa\\
$^{14}$Astrophysics Research Centre, School of Mathematics, Statistics, and Computer Science, University of KwaZulu-Natal, Westville Campus, Durban 4041, South Africa\\
$^{15}$Dunlap Institute of Astronomy \& Astrophysics, 50 St. George St., Toronto, ON M5S 3H4, Canada\\
$^{16}$David A. Dunlap Department of Astronomy \& Astrophysics, University of Toronto, 50 St. George St., Toronto, ON M5S 3H4, Canada\\
$^{17}$School of Earth and Space Exploration, Arizona State University, Tempe, AZ, USA 85287\\
$^{18}$Department of Physics and Astronomy, University of Pennsylvania, 209 South 33rd Street, Philadelphia, PA, USA 19104\\
$^{19}$A. Astrophysics Research Centre, University of KwaZulu-Natal, Westville Campus, Durban 4041, South Africa\\
$^{20}$B. School of Mathematics, Statistics \& Computer Science, University of KwaZulu-Natal, Westville Campus, Durban 4041, South Africa\\
$^{21}$Physics and Astronomy Department, Stony Brook University, Stony Brook, NY 11794\\
$^{22}$Instituto de F\'isica, Pontificia Universidad Cat\'olica de Valpara\'iso, Casilla 4059, Valpara\'iso, Chile\\
$^{23}$Joseph Henry Laboratories of Physics, Jadwin Hall, Princeton University, Princeton, NJ, USA 08544\\
$^{24}$McWilliams Center for Cosmology and Astrophysics, Department of Physics, Carnegie Mellon University, Pittsburgh, PA 15213, USA\\
$^{25}$Department of Physics, Cornell University, Ithaca, NY 14853, USA\\
}

\date{}

\pubyear{2024}

\begin{document}

\label{firstpage}
\pagerange{\pageref{firstpage}--\pageref{lastpage}}
\maketitle

\begin{abstract}
Patchy reionization generates kinematic Sunyaev-Zel’dovich (kSZ) anisotropies in the cosmic microwave background (CMB). Large-scale velocity perturbations along the line of sight modulate the small-scale kSZ power spectrum, leading to a trispectrum (or four-point function) in the CMB that depends on the physics of reionization. 
We investigate the challenges in detecting this trispectrum  and use tools developed for CMB lensing, such as realization-dependent bias subtraction and cross-correlation based estimators, to counter uncertainties in the instrumental noise and assumed CMB power spectrum.  We also find that both lensing and extragalactic foregrounds can impart larger trispectrum contributions than the reionization kSZ signal. We present a range of mitigation methods for both of these sources of contamination, validated on microwave-sky simulations.  We use ACT DR6 and \planck\ data to calculate an upper limit on the reionization kSZ trispectrum from a measurement dominated by foregrounds.  The upper limit is about 50 times the signal predicted from recent simulations.  
\end{abstract}




\begingroup
\let\clearpage\relax
\endgroup
\newpage

\section{Introduction}

The epoch of reionization (EoR) is the period during which the Universe's first stars and galaxies formed, and the resulting radiation 
began the process of ionizing the Universe, eventually leading to a Universe populated with the low-density, ionized hydrogen we observe today.
The EoR is a challenging period to observe, as the high-redshift ionizing sources are extremely faint, although recent data from the James Webb Space Telescope (JWST) presents a dramatic increase in available data on the formation of these first galaxies (e.g. \citealt{eisenstein23,finkelstein23a,finkelstein23b}). While current and upcoming 21cm experiments, such as HERA\footnote{\url{https://reionization.org/}} and SKA\footnote{\url{https://www.skao.int/en}}, aim to map out the neutral hydrogen abundance during reionization, they will  need to address foregrounds that are orders of magnitude above the signal.
    
Reionization also leaves imprints on the cosmic microwave background (CMB). As well as a large-scale polarization signature (e.g. \citealt{kogut03}), there is a contribution to the kinematic Sunyaev-Zeldovich (kSZ) signal from reionization, commonly referred to as the `patchy' kSZ effect (e.g. \citealt{santos03}, \citealt{mcquinn05}, \citealt{trac11}, \citealt{shaw12}, \citealt{battaglia13}). During reionization, there are large inhomogeneites in the free electron density as ``bubbles'' surrounding ionizing sources ionize before other regions. These bubbles will have a range of radial peculiar velocities with respect to the observer, leading to a scattering and Doppler shifting of the CMB photons that corresponds to a boosting (for bubbles moving towards us) or reduction (for bubbles moving away) in the observed CMB temperature. This generates a contribution to the kSZ power spectrum of the same order of magnitude as the low redshift effect, which arises due to radially moving inhomogeneities in the (now fully ionized) electron density field. 

The kSZ power spectrum dominates over the blackbody primary CMB signal at small scales, $l \gtrsim 4500$. Data from ground-based, high resolution, low noise, CMB experiments such as the South Pole Telescope (SPT) and the Atacama Cosmology Telescope (ACT, which we use here) already have greater sensitivity to this small-scale regime than \planck.
While there have been recent attempts to disentangle this reionization kSZ signal from the different contributions to the small-scale CMB temperature \citep{plankreion,reichardt21,choi20,gorce22} using only the power spectrum, the presence of foregrounds (including the late-time kSZ) has thus far posed significant challenges. These challenges are also discussed  in \citet{calabrese14} and \citet{sogoals} who forecast the potential of upcoming experiments like Simons Observatory\footnote{\url{https://simonsobservatory.org/}} to constrain reionization using the kSZ power spectrum. 

    
In this work, we focus on the non-Gaussian signature imparted on the CMB due to the kSZ effect during reionization. As laid out in \citet{smith17} and \citet{ferraro18}, this non-Gaussianity should be present due to the modulation of the kSZ signal by large scale radial velocity modes. For a given line of sight $\unitvec$, the small-scale kSZ power spectrum is amplified depending on the squared radial velocity field, projected along the line-of-sight, in direction $\unitvec$, with respect to the sky-averaged projected squared radial velocity. Given the large coherence length of the velocity perturbations, even after the line-of-sight projection there should be significant variation of the (squared) radial velocity across the sky, leading to a position-dependent kSZ power spectrum, i.e. a trispectrum. Both the power spectrum and the kSZ trispectrum include contributions from moving ionized overdensities in the more recent universe ($z \lesssim 3$) as well as from reionization. The 
trispectrum gives us additional information with which to disentangle the  contributions from the two kSZ epochs \citep{alvarez20}. 

While this paper was in collaboration review, \citet{raghunathan24} presented 
a first upper limit on the kSZ trispectrum using a combination of SPT and \textit{Herschel}-SPIRE data, finding that the contribution from foregrounds dominates over the kSZ signal. In this work  we also find that foregrounds present a major challenge for measuring the kSZ trispectrum, dominating over the predicted signal for current ACT + \planck\  data.  With upcoming, lower noise data however, we expect that residual foregrounds can be greatly reduced (because, e.g., fainter sources and clusters can be detected and mitigated against), and with the mitigation methods described in \sect{sec:fgmethods}, the kSZ trispectrum may become an informative probe of reionization physics.


Estimating an unbiased trispectrum from CMB data which has complex, anisotropic noise, and non-Gaussian foregrounds, has various challenges. In \sect{sec:theory} we describe our trispectrum estimator and bias corrections. In \sect{sec:fgmethods} we describe the biases due to lensing and foregrounds, and our mitigation strategies. In \sect{sec:fg_sims} we estimate foreground biases from microwave-sky simultions. In \sect{sec:data} we describe the ACT DR6+\planck\ data  and simulations used, and in \sect{sec:results} we describe our measurement. In \sect{sec:discussion} we conclude and discuss future improvements.

\section{Theoretical background and trispectrum estimator}\label{sec:theory}

In \sect{sec:est} we briefly outline and motivate the trispectrum estimator we use, and refer the reader to \citet{smith17} and \citet{ferraro18} for further details. We also discuss the sensitivity of the trispectrum statistic to the properties of reionization in \sect{sec:sens}, and discuss the bias corrections required in \sect{sec:n0andmf}.

\subsection{The Estimator}\label{sec:est}

The kSZ effect imparts a fractional temperature fluctuation on the CMB:
\begin{equation}
    \Delta \Theta^{\mathrm{kSZ}} (\unitvec) = -\frac{1}{c} \int \dx{\chi} g(\chi) \left[1+\delta_e(\chi\unitvec)\right] v_r(\chi\unitvec),
    \label{eq:tksz}
\end{equation}
where $\chi$ is the comoving distance, $g(\chi) = e^{-\tau}\frac{\dx{\tau}}{\dx{\chi}}$ is the visibility function, $\tau$ is the optical depth, $\delta_e(\chi\unitvec)$ is the free electron overdensity, and $v_r(\chi\unitvec) = \unitvec \cdot \bm{v}(\chi\unitvec)$ is the line-of-sight component of the peculiar velocity. As we will describe, in this work we are interested in the small-scale (high-$l$) temperature fluctuation generated by the kSZ, which depends on small-scale electron density perturbations, hence we drop the drop the constant electron density term in \eqn{eq:tksz}.  In the flat-sky limit, in the Fourier domain, the multiplication becomes a convolution:
\begin{equation}
\Delta \Theta(\vecl) = -\frac{1}{c} \int \dx{\chi} g(\chi) \int \frac{\dx{^2\vecl'}}{(2\pi)^2} \delta_e(\vecl'-\vecl, \chi) v_r(\vecl', \chi),
\end{equation}
where $v_r(\vecl, \chi) = \int d\Omega e^{i\vecl \cdot \unitvec} \unitvec \cdot \bm{v}(\chi\unitvec)$ and $d\Omega$ is the area element on the sky.

Ignoring for now the integral over the line-of-sight, it is useful to consider the case where the radial velocity field has only a single-large scale velocity mode $V_{\vecL}$, i.e. $v_r(\vecl', r) = \delta_D(\vecl'-\vecL)V_{\vecL}$. Ignoring other contributions to the temperature, we would then have 
\begin{align}
    \Delta\Theta(\vecl) &\propto \int \dx{^2\vecl}' \delta_e(\vecl-\vecl') \delta_D(\vecl'-\vecL)V_{\vecL}\\
    &\propto V_{\vecL} \delta_e(\vecl-\vecL)
\end{align}
where $\delta_D(\vecl)$ denotes the Dirac delta function. The product of two temperature modes is then
\begin{equation}
    \Delta\Theta(\vecl_1)\Delta\Theta(\vecl_2) \propto 
    V_{\vecL}^2 \delta_e(\vecl_1-\vecL) \delta_e(\vecl_2-\vecL) 
\end{equation}
and has expectation, for fixed $V_{\vecL}$ 
\begin{equation}
    \left< \Delta\Theta(\vecl_1)\Delta\Theta(\vecl_2)\right>_{\mathrm{fixed} V_{\vecL}} \propto V_{\vecL}^2 \delta_D(\vecl_1+\vecl_2-2\vecL) C_{\left|\vec{l}_1-\vec{L}\right|}^{ee},
\end{equation}
where $C_{l}^{ee}$ is the electron density power spectrum. We see then that a product of two temperature modes is an estimator for the  radial velocity squared $V_{\vecL}^2$, i.e. the kSZ effect enables us to probe the Universe's velocity field, a rich source of cosmological information, using the CMB.

The physical picture here is that the small-scale power in the CMB temperature due to the kSZ is modulated up or down for a given line-of-sight $\unitvec$, depending on the size of $\left< V_{\vecL}^2 \right>(\unitvec)$ for that region (where angle brackets imply an averaging along the line-of-sight). Note that in reality the kSZ effect has a broad redshift kernel, so the quadratic estimator is sensitive to the radial velocity squared projected over this broad redshift kernel. 
Given the large coherence length of the velocity field, there is still significant variation across the sky in this projected radial velocity squared, and this modulation generates a position-dependent power spectrum. Note that the trispectrum (i.e. four-point function) is the lowest-order non-zero statistic of the kSZ temperature perturbations  beyond the power-spectrum, since the bispectrum (i.e. three-point function) is zero by symmetry.\footnote{One can argue this in the following way: Consider the correlation of three radial velocity modes $\left< v_r(k_1) v_r(k_2) v_r(k_3)\right>$. The configuration with $v_r\rightarrow-v_r$ is equally likely, in which case the three-point function is $-\left< v_r(k_1) v_r(k_2) v_r(k_3)\right>$. In the ensemble average over possible values of $v_r$, these two configurations will always cancel, and $\left< v_r(k_1) v_r(k_2) v_r(k_3)\right>$, averaged over realizations of the $v_r$ field, is zero.}

Motivated by this, \citet{smith17} formed a quadratic estimator for this non-Gaussianity induced by the kSZ at reionization
\begin{equation}
    K_{\vecL} = 
    N^{KK}_{\vecL} \int \frac{\dx{^2\bm{l}}}{(2\pi)^2} W_{\vecl} W_{\vecL-\vecl} T_{\vecl}T_{\vecL-\vecl} \label{eq:KL}
\end{equation}
with the filter function,
\begin{equation}
W_{\vecl} = \sqrt{C_l^{\mathrm{pkSZ}}} / C_l^{\mathrm{tot}},
\end{equation}
chosen to up-weight the kSZ contribution to the temperature modes. Here $C_l^{\mathrm{pkSZ}}$ is the contribution to the temperature power spectrum due to the kSZ at reionization (i.e. the `patchy' term), and $C_l^{\mathrm{tot}}$ is the total observed temperature power spectrum (i.e. including noise). Throughout this work, when constructing the filter $W_l$, we set $C_l^{\mathrm{pksz}}$
equal to the angular power spectrum measured from the \citet{alvarez16} simulations, and use as $C_l^{\mathrm{tot}}$ a smoothed measurement of the power spectrum of the (beam-deconvolved) data temperature map (see \sect{sec:data} for more details).
In the squeezed limit $L\ll{l}$, $K$ is just a measure of the local amplitude of the ($l$-weighted) temperature power spectrum, integrated over scales $l$. One can think of $K(\unitvec)$ then as some measure of the local temperature power spectrum around the direction $\unitvec$. The power spectrum of $K$, $C_L^{KK}$, is a measure of how much it varies across the sky. Note that throughout, we will use $\vecL$ to denote the large scale on which we measure $K$, while $\vecl$ labels the (small scale) CMB temperature fluctuations which enter our quadratic estimator for $K_{\vecL}$.

$N_{\vecL}^{KK}$ is a normalisation factor which we define following \citet{sailer20}:
\begin{equation}
    (N_{\vecL}^{KK})^{-1} = \int\frac{\dx{^2\bm{l}}}{(2\pi)^2} 
    W_{\vecl}^2 W_{\vecL-\vecl}^2 \Cltot{\vecl}\Cltot{\vecL-\vecl}
\end{equation}
We note that this normalization is fairly arbitrary - so long as measurement and theoretical prediction are normalised consistently, the exact choice is not crucial. Nonetheless, we discuss alternative approaches, including that taken by \citet{raghunathan24}, in \app{app:norm}.


As a baseline theory model, we measure $C_L^{KK}$ from the \citet{alvarez16} simulations, using the same $W_{\vecl}$ as in the ACT DR6+\planck\ data measurement (see \sect{sec:results}). We use this measurement as a template to which we compare the data measurement and foreground biases, often reporting a fitted amplitude parameter that simply re-scales this template, which we call $\Aksz$. We note that there is considerable theoretical uncertainty in the $C_L^{KK}$ signal and that the AMBER simulations \citep{chen23,trac22} provide an updated treatment relative to \citet{alvarez16}. They also include a range of reionization scenarios. We consider these more extensive simulations in \app{app:amber}.

\subsection{Sensitivity to properties of reionization}\label{sec:sens}

\citet{smith17} and \citet{ferraro18} motivate the following theoretical model for $C_L^{KK}$:
\begin{equation}
C_L^{KK} \propto \int dz \frac{H(z)}{\chi^2(z)} \left[\frac{\mathrm{d}C_l^{\mathrm{pksz}}}{\mathrm{d}z} \right]^2 P_{\eta}\left[L/\chi(z), z\right]
\end{equation}
where $\chi(z)$ is the comoving distance to redshift $z$ and
\begin{equation}
    \eta(\unitvec,z) = v_r^2(\unitvec,z)/\left< v_r^2(z,\unitvec) \right>_{\unitvec} 
\end{equation}
is the squared radial velocity contrast, i.e. the squared radial velocity divided by its sky average.
The dependence on the details of reionization enters largely through  $\left[\frac{\mathrm{d}C_l^{\mathrm{pksz}}}{\mathrm{d}z} \right]^2$, which is the derivative of the reonization contribution to the kSZ power spectrum as a function of redshift. Thus $C_L^{KK}$ may be increased by physics that increases $C_l^{\mathrm{pkSZ}}$, such as in models where more of the reionization occurs at higher redshift, since the physical electron density is higher (scaling as $(1+z)^3$). 
The trispectrum signal additionally depends on $P_{\eta}\left[L/\chi(z), z\right]$, the power spectrum of the squared radial velocity contrast. This largely sets the scale-dependence of the $C_L^{KK}$, with earlier reionization corresponding to the signal peaking on smaller (angular) scales. 

As discussed in \citet{smith17,ferraro18,alvarez20,raghunathan24} the kSZ trispectrum signal is mainly sensitive to the midpoint, $\zmid$ and width, $\Delta z$ of reionization. In particular, when more of the reionization occurs at higher redshift (which could result from larger $\zmid$ or $\Delta z$), the kSZ power spectrum tends to be increased since ionized bubbles of a given size have higher mean density, and thus higher optical depth. Since the trispectrum  signal arises from the modulation of this power spectrum, it also increases. For the trispectrum signal we note that there should be a competing effect whereby larger $\Delta z$ means the modulating radial velocity field is averaged over a longer line-of-sight and therefore should have smaller fluctuations. Hence we do expect a somewhat different sensitivity to changes in the reionization parameters for the power spectrum and the trispectrum, allowing for breaking of the $\tau-\Delta z$ degeneracy demonstrated in \citet{alvarez20}. 

The complementary information provided by the trispectrum is likely even more important when realistic foregrounds are present, making interpretation of either statistic alone more difficult.
In \app{app:amber} we study the response of the reionization kSZ power spectrum and trispectrum to changes in the reionization parameters.

\subsection{$N^0$ bias, mean-field and the cross-correlation estimator}\label{sec:n0andmf}

There is a bias to $C_L^{KK}$ arising from the disconnected trispectrum, i.e. the trispectrum one would measure for a Gaussian field with the same power spectrum. 
We follow the CMB lensing convention here and denote this bias as $N^0_{\vecL}$. 

As in the case of CMB lensing, this correction depends on the total power spectrum of the input temperature map (i.e. including CMB, foregrounds and instrumental noise), and inaccuracy in the assumed total power spectrum will lead to bias in the calculated $N^0$ correction, thus biasing the inferred $C_L^{KK}$. Hence a  calculation where the estimator is simply applied to many Gaussian simulations with representative power, and averaged (an approach we will refer to as MC$N^0$), is susceptible to biases in that assumed representative power spectrum. As we will see, the $N^0$ bias can be  orders of magnitude larger than the kSZ signal we are trying to recover here, so even a small fractional bias in the $N^0$ can cause biases in $C_L^{KK}$.

In this work, we propose to follow the approach taken in CMB lensing and use a realization-dependent $N^0$ (RD$N^0$) \citep{hanson11, namikawa13} 
. This method removes biases in the $N^0$ that depend linearly on the fractional difference, $\Delta C_l / C_l^{\mathrm{tot}}$, between the true power spectrum and that assumed for the simulations. With this method remaining biases are then $\mathcal{O}\left[(\Delta C_l / C_l^{\mathrm{tot}})^2\right]$ or higher, thus a sub-percent accuracy $N^0$ can be achieved even with few-percent uncertainty in the total power spectrum.
We show in \sect{sec:results} that this method accounts for small errors in the assumed temperature power spectrum. We believe this is especially important given the high $l$s considered for this measurement ($l\gtrsim 3000$), where highly uncertain foregrounds and the kSZ signal itself dominate over the better understood primary CMB power spectrum.  

We also follow the lead of CMB lensing approaches by using the cross-correlation-only trispectrum estimator of \citet{madhavacheril20}, as described in \sect{sec:results}, which makes the $N^0$ bias independent of the instrumental noise (which would otherwise be difficult to model  accurately).
The cross-correlation-only estimator also removes the sensitivity to instrumental noise of the \emph{mean-field}, the bias that arises due to the statistical anisotropy of the survey mask and noise. We note that the cross-correlation estimator can result in reducing the $S/N$ of the measurement, since it removes the auto-correlations of the independent data splits. This reduction is significant in the high $l$ regime used here, where noise that is uncorrelated between data splits becomes dominant over signal. \citet{madhavacheril20} show that using a greater number of splits reduces the $S/N$ cost of the cross-correlation estimator. In this work we use four splits, which results in roughly a factor of 2 to 3 increase of our measurement uncertainties. As we will demonstrate in \sect{sec:results}, given that our upper-limit is dominated by foreground contamination rather than statistical uncertainty, we think the extra robustness provided by the cross-correlation estimator is worth the $S/N$ cost.

The $N^0$ and mean-field biases are thus accounted for in our estimator for $C_L^{KK}$:

\begin{multline}\label{eq:clkkcorr}
    C_L^{KK} = \left[ \left< (\hat{K}_{AB} - \bar{K}_{AB})_{\vecL}(\hat{K}_{CD}^* - \bar{K}_{CD}^*)_{\vecL} \right> - \right. \\ \left. \mathrm{RD}N^{0,ABCD}_{\vecL} \right] / (w_4 T_L^{KK}).
\end{multline}
In this expression the angled brackets, $\left<XY\right>$ denote the angular power spectrum between fields $X$ and $Y$. $\hat{K}^{AB}_{\vecL}$ is the ``raw'' (i.e. before mean-field correction) data measurement of the $K_{\vecL}$ field in \eqn{eq:KL}, computed from two temperature maps labelled $A$ and $B$. These two temperature maps may be the same or different (for example, if different foreground cleaning measures have been applied to each, see \sect{sec:freqclean} for discussion). $\bar{K}^{AB}_{\vecL}$ is the statistical anisotropy due to the noise and the survey mask, known in the CMB lensing literature as the \emph{mean-field}. It is calculated as the mean of the $K_L$ estimator applied to  Gaussian simulations which have the same statistical anisotropy due to noise and mask, but no non-Gaussian kSZ signal. $\mathrm{RD}N^{0,ABCD}_{\vecL}$ is the realization-dependent $N^0$ correction calculated using these same survey simulations. Finally, we also include multiplicative factors $w_4$\footnote{$w_4 = \sum_{i} a_i(m^i)^4 / \sum_{i} a_i$ where the sum is over pixels $i$, $a_i$ is the area of pixel $i$, and $m^i$ is the value of the (apodized)  mask (between 0 and 1) for pixel $i$.} and $T_L^{KK}$ (detailed in \sect{sec:bias_sims}) to account for the survey mask and survey transfer function. 

It is this estimator that we apply to ACT DR6 and \planck\ in \sect{sec:results}, where we also demonstrate the importance of using the cross-correlation-based estimator, and RD$N^0$, for our measurement. First though, we turn to the significant contamination of the $C_L^{KK}$ signal  by extra-galactic foregrounds and CMB lensing.

\section{New methods for mitigation of lensing and Foreground Biases}\label{sec:fgmethods}

Our trispectrum statistic, $C_L^{KK}$ is susceptible to contamination from other sources of non-Gaussianity in the observed CMB, in particular, gravitational lensing and extra-galactic foregrounds. In \sect{sec:fg_sims} we estimate biases to $C_L^{KK}$ using microwave sky simulations. In this section, we introduce and motivate methods to mitigate against these biases: \emph{lensing-hardening} and frequency-based methods. Note also that the use of RD$N^0$, as proposed in the previous section, will also mitigate the impact of foreground uncertainty via reducing the sensitivity to the assumed foreground contribution to the assumed power spectrum.

\subsection{Lensing and bias-hardening}\label{sec:bh}

Lensing of the CMB by a lensing potential field $\phi_{\bm{L}}$ generates a mode-coupling
\begin{equation}
    \left< \Theta(\vecl)\Theta(\vecL-\vecl) \right> =  f^{\phi}_{\bm{l},\bm{L}}\phi_{\bm{L}}
\end{equation}
where
\begin{equation} 
    f^{\phi}_{\bm{l},\bm{L}} = C_{l}\bm{l} \cdot \bm{L}+C_{|\bm{L}-\bm{l}|}\bm{l} \cdot (\bm{L}-\bm{l}).\label{eq:flL}
\end{equation}
and $C_l$ is the CMB power spectrum without noise.

Lensing therefore produces a bias to our $K$ estimate, $K^{\phi}_{\vecL}$, which has expectation 
\begin{equation}
    K^{\phi}_{\vecL} =  
    N^{K}_{\vecL} \phi_{\vecL} \int \frac{\dx{^2\bm{l}}}{(2\pi)^2} W_{\vecl} W_{\vecL-\vecl} f^{\phi}_{\bm{l},\bm{L}}.
\end{equation}
As we will show below, this signal is larger than the expected signal from the kSZ, but can be mitigated via bias-hardening, 
a method developed for CMB lensing \citep{namikawa13, osborne14} that aims to isolate the lensing potential from other sources of mode-coupling in the reconstruction, such as Poisson distributed point-sources or anisotropy due to the mask. When the functional form of mode-coupling is given, e.g. \eqn{eq:flL} for lensing, one can write down a response of some other quadratic estimator to that source of mode-coupling. Our case is most similar to that investigated in \citet{sailer20} 
, which used a bias-hardened estimator to remove the contamination to lensing due to Poisson-distributed extended sources. Noting the similarity of our $K$ estimator with the extended source estimator of \citet{sailer20}, here we can instead use bias-hardening to remove the contamination due to lensing from our $K$ estimator.


In general a bias-hardened estimator for a field $x_{\vecL}$ in the presence of a contaminant $y_{\vecL}$ is 
\begin{equation}
    x_{\vecL} = \frac{\hat{x}_{\vecL}-N^x_{\vecL} R^{xy}_{\vecL}\hat{y}_{\vecL}}{1-N^x_{\vecL}N^{y}_{\vecL} (R^{xy}_{\vecL})^2}
\end{equation}
where $\hat{x}_{\vecL}$ is the non-hardened estimator,
\begin{equation}
    R_{\vecL}^{xy} = \int \frac{\dx{^2\bm{l}}}{(2\pi)^2} \frac{f^{y}_{\bm{l},\bm{L}} f^{x}_{\bm{l},\bm{L}}}{2\Cltot{l}\Cltot{|\vecL-\vecl|}}
\end{equation}
and $N^{x}_{\vecL}$ is the estimator normalization, given by $1/R_{\vecL}^{xx}$. The functions $f^x_{\vecl,\vecL}$ and $f^y_{\vecl,\vecL}$ set the mode coupling generated by the fields $x$ and $y$, e.g. the function for lensing is given above in \eqn{eq:flL}.

The noise on the bias-hardened estimator is 
\begin{equation}
    N^{BH,x}_{\vecL} = N^{x}_{\vecL}/(1-N^x_{\vecL}N^y_{\vecL} (R^{xy}_{\vecL})^2).
\end{equation}

For the case of lensing, the contaminant field is $\phi_{\vecL}$, with $f^{\phi}_{\vecl,\vecL}$ given by \eqn{eq:flL}. Our `lensing-hardened' estimator is then
\begin{equation}
    K^{\mathrm{LH}}_{\vecL} = \frac{\hat{K}_{\vecL}-N^K_{\vecL} R^{K\phi}_{\vecL}\hat{\phi}_{\vecL}}{1-N^K_{\vecL}N^{\phi}_{\vecL} (R^{K\phi}_{\vecL})^2}.
    \label{eq:K_lh}
\end{equation}
We have implcitly assumed here that the mode coupling generated by the $K$ field is equivalent to that generated by Poisson distributed extended sources with profile $\sqrt{C_l^{\mathrm{kSZ}}}$, i.e. setting $f^{K}_{\vecl,\vecL}=\sqrt{C_l^{\mathrm{pkSZ}}C_{|\vecL-\vecl|}^{\mathrm{pkSZ}}}$. While this is an approximation, we only assume this in order to construct the lensing-hardened estimator, and test the accuracy of this below.


In \fig{fig:fg1} we show the lensing bias measured as the average of our $C_L^{KK}$ estimator applied to simulations of the lensed CMB. We use the same filtering and $l$-ranges as for our baseline data measurement (described in \sect{sec:results}). We find that lensing-hardening alone reduces the lensing bias to $C_L^{KK}$ by roughly an order of magnitude. We see also that the lensing-hardened estimator can be further improved by subtracting a simulation-based $N^1$ correction. This term, identified for the CMB lensing power spectrum estimator by \citet{kesden03}, is required to correct the $\left<\hat{\phi} \hat{\phi}\right>$ terms appearing in  $C_L^{KK}$ when the lensing-hardened estimator of \eqn{eq:K_lh} is used. 

Bias-hardening can also be used to remove the contribution from Poisson distributed point-sources (this was one of its  original motivations in the CMB-lensing context). We might expect this to significantly help with contamination from the cosmic infrared background (CIB) and radio galaxies. However, one would also expect a significant noise cost, since the form of mode-coupling induced is similar to our signal of interest (which appear roughly like Poisson distributed extended sources).

\subsection{Extra-galactic foregrounds and frequency-cleaning methods}\label{sec:freqclean}

Non-Gaussian signatures in the CMB are also generated by astrophysical sources of microwave radiation, such as radio sources and the CIB and the late-time kSZ and thermal SZ (tSZ) effects, which arise from inverse-Compton scattering of CMB photons. These will in general impart a trispectrum.  As well as contributions from the trispectrum of each individual foreground, various other terms will arise due to correlations between the foreground fields (e.g. the tSZ and CIB fields are quite correlated). 

Various strategies can be employed to reduce the contamination of the CMB temperature by foregrounds. Finding (e.g. via matched filter methods) and masking (or modeling, or inpainting) point-sources and clusters reduces the contribution from radio sources, the CIB and the tSZ. In addition, one can use the frequency spectral dependence of extra-galactic foregrounds to ``clean'' temperature maps. In internal linear combination (ILC) methods (see e.g. \citealt{remazeilles11}), linear combinations of observed frequencies are constructed which preserve the CMB signal, but null (or ``deproject'') one or more assumed foreground frequency spectra. We note that the contribution from the low-redshift kSZ can not be deprojected, since it has the same frequency spectral dependence as the reionization kSZ that we are aiming to measure.
Deprojecting foreground spectra will result in a noisier ILC map compared to the ILC where the only constraint is that the variance is minimized. For our case this noise cost is amplified by the fact that we are measuring a trispectrum of the temperature map (e.g. a factor of 2 increase in noise at the map level corresponds to a factor of 8 increase in noise in the trispectrum). 

However, here we are interested in the kSZ signal at reionization, which we can assume is to a good approximation uncorrelated with the lower redshift large-scale structure generating the foreground contamination. Unlike for the CMB lensing case then, to remove the foreground contribution from the trispectrum, we need not use a foreground-cleaned map in all four legs of the $C_L^{KK}$ estimator. Denoting our $K$ estimator from two temperature maps $\Theta_A$ and $\Theta_B$ as $K(\Theta_A,\Theta_B)$, then the following $C_L^{KK}$ estimators are all unbiased by foregrounds:
\begin{align}
    &\left< K(\Theta_{\mathrm{clean}}, \Theta_{\mathrm{clean}}), K(\Theta_{\mathrm{clean}}, \Theta) \right>, \\ 
    &\left< K(\Theta_{\mathrm{clean}}, \Theta), K(\Theta_{\mathrm{clean}}, \Theta) \right>, \\
    &\mathrm{and} \left< K(\Theta_{\mathrm{clean}}, \Theta), K(\Theta, \Theta )\right>,
\end{align}
where $\Theta_{\mathrm{clean}}$ is a foreground-cleaned map, and $\Theta$ is the minimum-variance ILC map (see e.g. \citealt{madhavacheril18,darwish21} for similar approaches, sometimes referred to as `gradient-cleaning', in CMB lensing and \citealt{kusiak21} for a similar approach for the projected fields kSZ estimator).

These estimators will however have different noise properties, since the statistical uncertainty on the trispectrum depends on the power spectrum of (and level of correlation between) the input maps. In the realistic case that foreground cleaning does not work perfectly (e.g. since the CIB is not perfectly described by a single $l$-dependent frequency spectrum), they will also have different levels of residual foreground bias. 

We will explore the performance of various of these estimators on simulations in \sect{sec:fg_sims}.

\section{Foreground contamination estimates}
\label{sec:fg_sims}

In order to predict the levels of extra-galactic foreground contamination, we use the ACT DR6-like versions of the \websky\ \citep{websky} and \citet{sehgal10} (\sehgal\ henceforth) simulations produced in \citet{maccrann23} -- see Section 4 of that work for a description of the simulation processing. The simulations  include tSZ, late-time kSZ, radio sources and the CIB. After adding simulated CMB and realistic ACT DR6-like noise, we run the \nemo\footnote{\url{https://nemo-sz.readthedocs.io/en/latest/}}\ source and cluster finding code, and use the outputs to subtract models for $S/N>4$ point sources, and the tSZ contamination from $S/N>5$ clusters (the same thresholds applied in the ACT DR6 data processing). The predicted contamination to $C_L^{KK}$, which we will refer to as $\Delta C_L^{KK}$, is simply estimated by measuring the $C_L^{KK}$ of the total (i.e. summed over all astrophysical components) residual foreground map. We correct for the mean-field (necessary since we apply a mask to the foreground simulations in order to make realistic the source and cluster finding) and $N^0$.\footnote{For the bias-hardened case, this is given by 
\begin{equation}
    N^{0}_{\mathrm{fg,BH}} = \frac{
    N^0_{\mathrm{fg}} + (R^{xy})^2 (N^{x})^2 N^{y} - 2 R^{xy} N^{y} (A^{x})^2 R^{xy}_{\mathrm{fg}} }{\left[1 - N^x N^y (R^{xy})^2\right]^2},
\end{equation}
where $N^0_{\mathrm{fg}}$ is the $N^0$ for the foreground-only map without bias hardening, and all quantities are functions of $L$.
}
%
%

As described in \sect{sec:freqclean}, in each leg of our trispectrum estimator, we can use a (possibly different) linear combination of the temperature maps from our range of frequency channels (channels centered at 97 and 149 GHz from ACT, and the 217, 353 and 545 GHz channels from \planck). We test linear combinations that minimize the variance in harmonic space, and also the case where frequency spectra for tSZ and CIB are additionally deprojected. Specifically, we generate temperature maps with the following configurations:
\begin{itemize}
    \item Minimum variance ILC, ``MV ILC'' henceforth.
    \item tSZ and CIB-deprojected ILC, ``tSZ+CIB-nulled ILC'' henceforth. We assume a CIB frequency spectrum:
\begin{equation}
    f_{\mathrm{CIB}}(\nu) \propto 
    \frac{\nu^{3+\beta}}{e^{h\nu/k_B\Tcib}-1}
    \left(\left.\dfrac{\dx{B}(\nu,T)}{\dx{T}}\right\vert_{T_{\mathrm{CMB}}}\right)^{-1}
    \label{eq:cib_spec}
\end{equation}
where $h$ is Planck's constant, $k_B$ is the Boltzmann constant, $B(\nu,T)$ is the Planck function and $T_{\mathrm{CMB}}$ is the CMB temperature.
We set $\beta = 1.2$ and $\Tcib=24K$ (following \citealt{sogoals}).  We note this is only an approximation to the true ($l$-dependent) CIB frequency spectrum, so we should not expect CIB to be perfectly removed.

\end{itemize}
We also investigated CIB-deprojected ILC (without tSZ-deprojection), as well as deprojecting additional components using the `moments method' \citep{chluba17,rotti21}, but find that they do not reduce biases further (we think that given the relatively high noise in the high frequency \planck\ channels at the high $l$ we are using, there is insufficient information to usefully constrain the additional components introduced in the moments method).

\begin{figure*}
    \centering
    \includegraphics[width=0.98\textwidth]{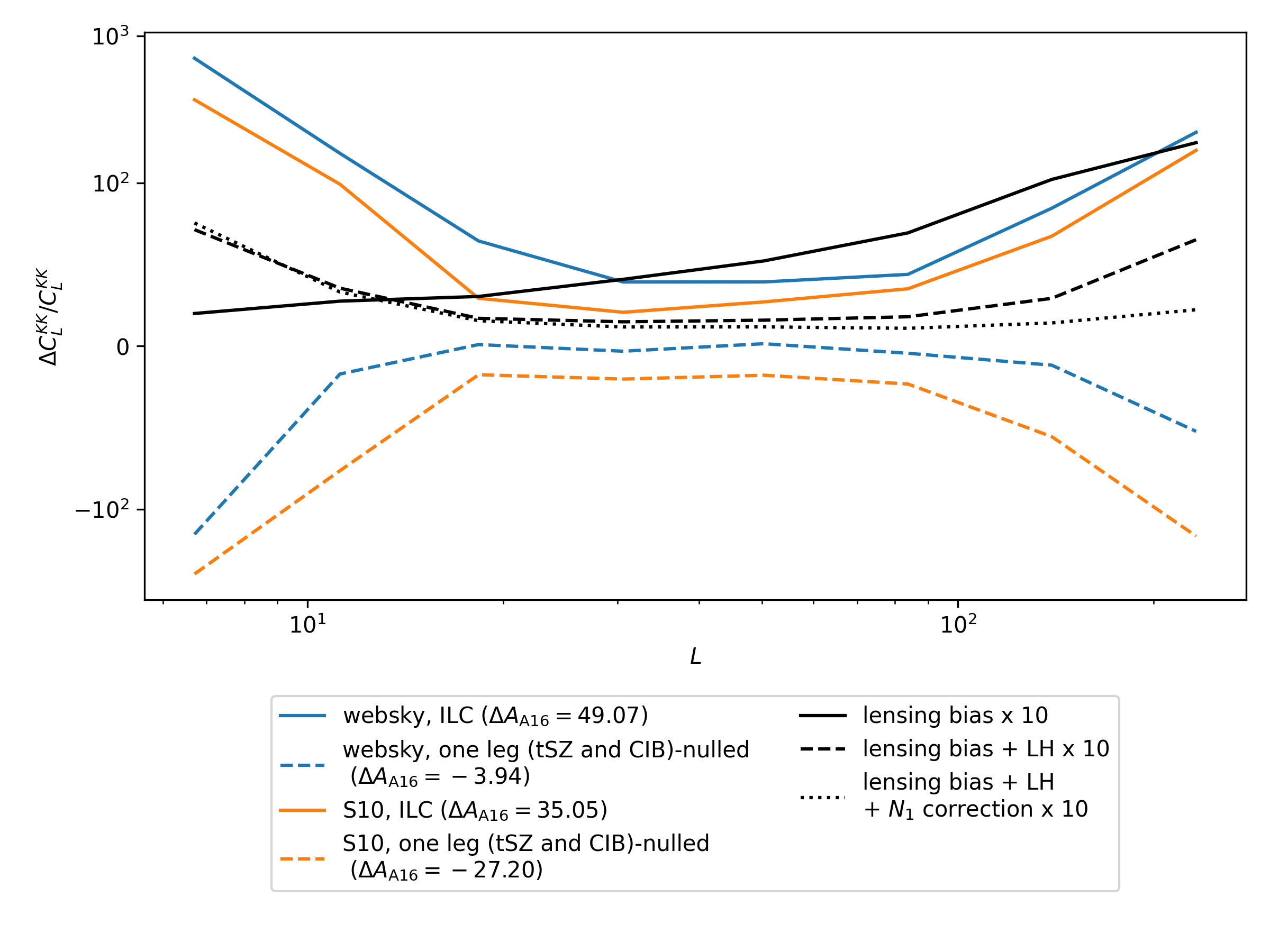}
    \caption{Expected fractional foreground biases from the \websky\ (blue lines) and \sehgal\ (orange lines) simulations, for our baseline CMB temperature $l$ range $3000< l < 4000$. Note the y-axis is linear in the range [-100,100], and log-scaled otherwise. Solid lines use a minimum variance ILC to combine frequency channels. For dashed lines, one of the temperature maps in the trispectrum has tSZ and CIB spectra de-projected (see \sect{sec:freqclean}). In the legend, we quote $\Delta \Aksz$ which is the bias in the amplitude fitted to the $C_L^{KK}$ prediction from the \citet{alvarez16} simulations. We also show the bias due to lensing multiplied by 10, without lensing mitigation (black solid line),  when lensing-hardening is used  (``LH'', dashed black line, see \sect{sec:bh}) and when an additional $N^1$ correction is used (dotted black line). 
    }
    \label{fig:fg1}
\end{figure*}

We show in \fig{fig:fg1} the fractional bias to $C_L^{KK}$ predicted from the \websky\ (blue lines) and \sehgal\ (orange lines) simulations for several estimator variations. The bias, $\Delta C_L^{KK}$ is divided by the \citet{alvarez16} $C_L^{KK}$ prediction, hence can be interpreted as a fractional bias if one assumes the \citet{alvarez16} prediction is correct. In the legend we also include the predicted bias to $\Aksz$, denoted $\Delta \Aksz$. We show the bias for the following cases, which we find to perform best:
\begin{enumerate}
    \item We use a minimum-variance ILC in all trispectrum legs, labelled ``ILC''.
    \item the same as (i) but replacing the temperature map in just one leg with a tSZ and CIB-nulled version, labelled `one leg (tSZ and CIB)-nulled'.
\end{enumerate}
We also checked the case where we deproject tSZ and CIB from more than one trispectrum leg, but this generally results in larger foreground biases\footnote{this can occur when residual, i.e. post-deprojection foreground contributions are amplified by the large weights required to null the assumed frequency spectra in the ILC.}, as well as higher noise.
While the `ILC' case shows very significant biases for both simulations, this estimator does have the advantage that the foreground biases are very likely positive\footnote{This is true for Poisson distributed foreground sources, which is likely a good approximation for the high CMB temperature $l$s we use here. We note also that in both \websky\ and \sehgal\ simulations the foreground bias for the ILC case is strongly positive.} 
, so a data measurement can be interpreted simply as an upper-limit on the reionization kSZ trispectrum. In contrast, the case where one trispectrum leg is `cleaned' may generate positive or negative foreground biases (since the residual foreground contribution in the cleaned leg may be correlated or anti-correlated with the foreground contributions in the other legs). While there is still significant uncertainty in the theoretical prediction of the foreground bias (as evidenced by the difference in the \websky\ and \sehgal\ predictions), we conclude that the strict positive upper-limit makes the ILC case more interpretable, and we use that as our baseline case. 

We note here that for our current data, the biases due to extragalactic foregrounds are significantly larger than those due to lensing (especially when lensing-hardening is used) - note that the lensing biases have been multiplied by 10 in \fig{fig:fg1}. There will also exist biases to $C_L^{KK}$ arising from the correlation between lensing and extragalactic foregounds, i.e. trispectrum terms like 
\begin{align}
&\left< K(\Theta_{\mathrm{cmb}}, \Theta_{\mathrm{cmb}}) K(\Theta_{\mathrm{fg}}, \Theta_{\mathrm{fg}})\right> \\ 
\mathrm{and} &\left< K(\Theta_{\mathrm{cmb}}, \Theta_{\mathrm{fg}}) K(\Theta_{\mathrm{cmb}}, \Theta_{\mathrm{fg}}) \right>
\end{align}
which depend on the bispectrum $\left< \phi \Theta_{\mathrm{fg}} \Theta_{\mathrm{fg}} \right>$, where $\Theta_{\mathrm{fg}}$ is the temperature perturbation due to foregrounds. However, given the dominance of the foreground-only biases for our current data, we do not consider those cross-terms here.



We investigate in \app{sec:fgmax} the dependence of the fractional foreground bias on the maximum CMB temperature multipole, $l$ used, for the ILC case. For both \websky\ and \sehgal, the foreground biases increase when increasing $\lmax$, and we decide to use $l_{\mathrm{max}}=4000$ as our baseline choice. We use $\lmin=3000$ throughout. We conclude that extragalactic foregrounds are a major systematic for the kSZ trispectrum, and will dominate the limits on the kSZ reioniation trispectrum in this  work.

\section{ACT DR6 + \planck\ data and simulations used}\label{sec:data}

We use CMB data from the Atacama Cosmology Telescope (ACT). ACT was a telescope located in the Atacama Desert in northern Chile and observed the millimetre wavelength sky from 2007 until 2022. For this analysis we 
use night-time temperature data collected from 2017 to 2022 in 
the CMB-dominated bands \texttt{f090} (77-112 GHz) and \texttt{f150} (124-172 GHz). These maps have a pixel scale of $0.5^\prime$ resolution, with a beam full-width-half-maximum of 2.2 and 1.5 arcminutes for  \texttt{f090} and \texttt{f150} respectively.  These  observations at \texttt{f090} (\texttt{f150}) were made using two (three) dichroic detector modules, known as polarization arrays (PAs) \citep{thornton16}. For each frequency channel and PA, four independent data split maps were constructed by using different subsets of the time-ordered data.
For this analysis, we obtain 4 independent data splits for a given frequency channel by combining the data for given split from each PA. We do this by performing an inverse-variance coadd in harmonic space (following \citealt{qu23}, see Section 5.6 of that work for details). 
allowing us to use the cross-correlation estimator described in \sect{sec:n0andmf}. 

We find approximate equivalent white noise levels of 12.4 $\mu$K-arcmin and 16.3 $\mu$K-arcmin for \texttt{f090} and \texttt{f150} respectively, for the coadd of the four independent splits.
We subtract and inpaint a catalogue of 1779 objects that include especially bright sources and extended sources with SNR$>10$. We further subtract tSZ model images corresponding to clusters detected with the \texttt{NEMO} software.  Refer to \citet{maccrann23} for further details regarding point-source and cluster template subtraction.
We also include the \planck\ \textsc{npipe} \citep{plancknpipe} 217, 353 and 545 GHz channels, from which we model and subtract $S/N>5$ point sources. 

For all frequencies (ACT and \planck) we mask Fourier modes with $|\ell_x|<90$ and $|\ell_y|<50$ to remove contamination  by ground, magnetic, and other types of pick-up in the data due to the scanning of the ACT telescope. 
We apply a $60\%$ galactic sky mask, with a 3 degree apodization, following \citet{qu23}.


For \planck\ we do not have four independent data splits, so we just use the same \planck\ data in each leg of the trispectrum estimator. This means our mean-field and $N^0$ is independent of ACT instrumental noise, but not of \planck\ instrumental noise. However, the \planck\ noise is easier to simulate (e.g., as there is no atmospheric noise and the scan strategy is simpler). Furthermore, for our baseline result where we use a minimum-variance ILC, the \planck\ channels contribute very little weight to the ILC anyway given their higher map noise and larger beam.

We construct the covariance matrix for harmonic ILC weights using heavily-smoothed\footnote{we use a Savitzky-Golay filter \citep{Savitzky64} of window length 301, order 2, as implemented at \url{https://docs.scipy.org/doc/scipy/reference/generated/scipy.signal.savgol_filter.html}} auto and cross-power spectra measured from each ACT and \planck\ frequency. For the ACT channels, we measure these power spectra from the average of the four independent data splits. Having constructed the ILC maps, we again measure heavily-smoothed, beam-deconvolved angular power spectra, which are used as $C_l^{\mathrm{tot}}$ in the filter function $W_l = \sqrt{C_l^{\mathrm{pkSZ}}}/C_l^{\mathrm{tot}}$.

We did not find significant improvement from using the Needlet ILC method presented in \citet{coulton24}, perhaps because in the high $l$ regime considered here, the foregrounds are fairly isotropic. Hence we decided to stick with a harmonic ILC approach given the lesser computational cost.

\subsection{Simulations for bias corrections and covariance}\label{sec:bias_sims}

To accurately calculate the RD$N^0$ and mean-field bias corrections we require simulations with the same power spectrum as the data. For the RD$N^0$ calculation the simulations need not contain a non-Gaussian kSZ component 
, while for the mean-field calculation Gaussian signal simulations should also suffice, since we are interested only in the statistical anisotropy generated by the noise and the mask. 
We generate 200 Gaussian signal simulations, ensuring these have the correct power spectrum via the following procedure:
\begin{enumerate}
    \item{Measure auto and cross-power spectra for all frequency (or ILC) maps, apply a $w_2$ correction to account for the mask\footnote{we divide all power spectra by $w_2 = \sum_{i} a_i(m^i)^2 / \sum_{i} a_i$ where the sum is over pixels $i$, $a_i$ is the area of pixel $i$, and $m^i$ is the value of the (apodized)  mask (between 0 and 1) for pixel $i$}, and apply a smoothing filter\footnote{we use a Savitzky-Golay filter of window length 101, order 2}. Call this $C_{l,ij}^{\mathrm{data}}$ for frequencies $i$ and $j$. We checked our results are not sensitive to using an increased smoothing window length.}
    \item{Generate Gaussian random fields with covariance $C_{l,ij}^{\mathrm{data}}$. Apply the survey mask and k-space masking. Measure the resulting power spectra $C_{l,ij}^{\mathrm{cal}}$, again applying the $w_2$ correction.}
    \item{Define 
    \begin{equation}
        T_{i} = C_{l,ij}^{\mathrm{cal}}/C_{l,ij}^{\mathrm{data}}
    \end{equation}}
    \item{Generate Gaussian random fields with power $C_{l,ij} = C_{l,ij}^{\mathrm{data}} / \sqrt{T_i T_j}$. Apply survey mask and k-space masking as on the data.}
\end{enumerate}

We then add these signal simulations to noise-only simulations generated following \citet{atkins23} for the ACT channels, and using those provided by \citet{plancknpipe} for the \planck\ channels.
We use these  simulations to calculate the mean-field and RDN0 biases, from which we can calculate our corrected $C_L^{KK}$ according to \eqn{eq:clkkcorr}. 

To compute the covariance matrix and $T_L^{KK}$ (see \eqn{eq:clkkcorr}), we add a randomly rotated version of the \citet{alvarez16} reionization kSZ simulation to each realisation. The covariance matrix is then computed from the application of our $C_L^{KK}$ estimator (\eqn{eq:clkkcorr}) to each of these simulation realisations. We do not re-calculate RD$N^0$ for every simulation realisation, instead replacing it with MC$N^0$ (see \sect{sec:n0andmf}). This likely results in a small over-estimation of the off-diagonal covariance, however, we will show that our upper limit on the kSZ trispectrum is dominated by the foreground bias rather than statistical uncertainty, so we believe it is reasonable to avert the very high computational cost of including the RD$N^0$ correction for every realisation. 
We note also that the covariance matrix simulations do not then include non-Gaussian foregrounds, which may result in an underestimate of the covariance. Again, given that the the upper limit on the kSZ trispectrum is dominated by 
foreground bias rather than statistical uncertainty, we do not believe this potential underestimate significantly affects our conclusions. 

Finally, we use these simulations to compute a multiplicative correction to $C_L^{KK}$ that arises largely from the Fourier mode masking described in \sect{sec:data}. For each simulation realisation we compute the ratio of our mean-field, $N^0$ and $w_4$-corrected $C_L^{KK}$ estimator to the input, reionization kSZ-only $C_L^{KK}$. Averaged over realisations, we infer a close to scale-independent multiplicative bias  $T_L^{KK}=0.94$.

\section{Results}\label{sec:results}

\subsection{ACT DR6 + \planck\ Measurements of $C_L^{KK}$}\label{sec:meas}

\begin{figure*}
    \includegraphics[width=0.8\textwidth]{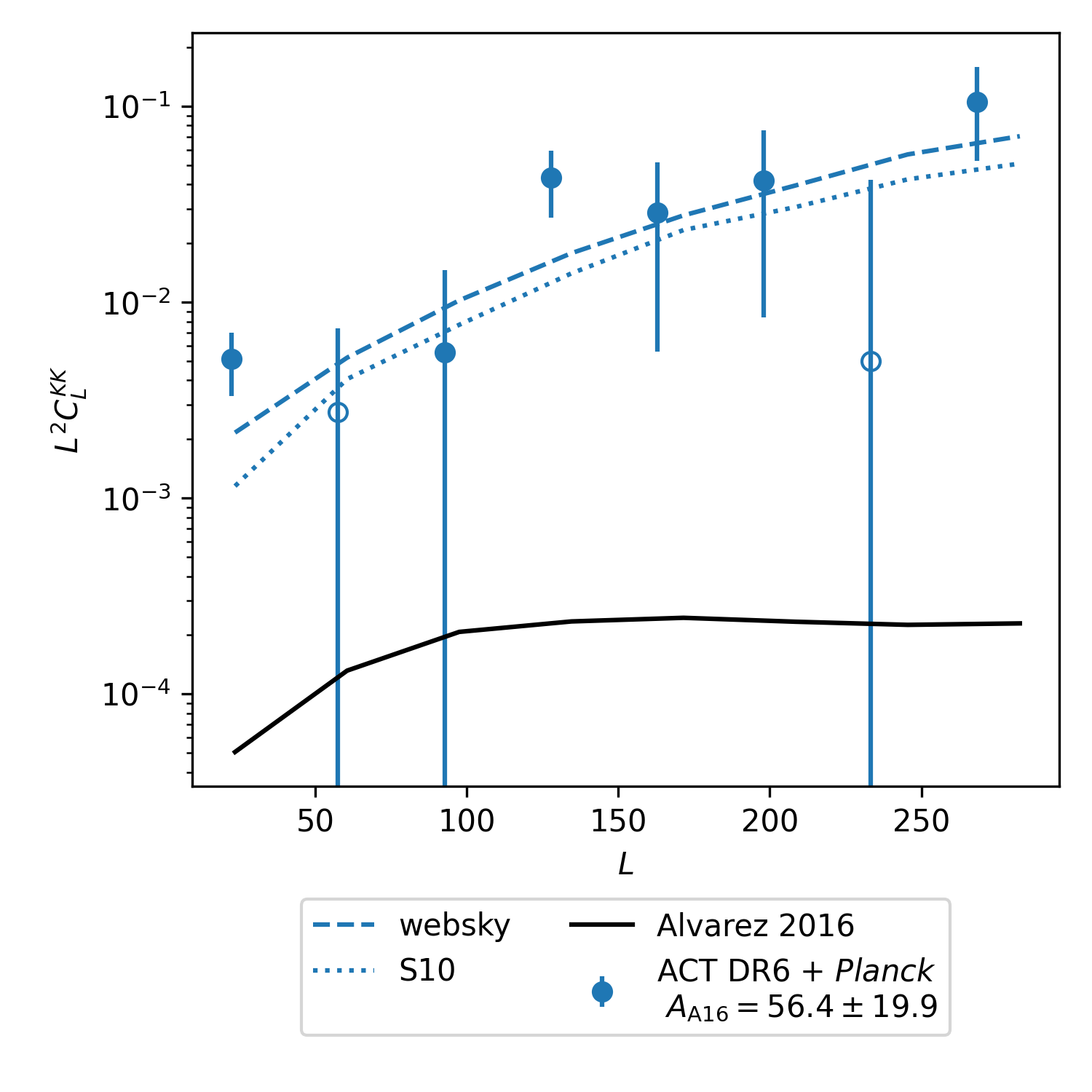}
    \caption{
    Blue points show $C_L^{KK}$ from ACT and \planck\ data (combined via a harmonic space ILC). Negative points are indicated by open symbols. The dashed and solid lines show the corresponding predictions of the foreground bias from \websky\ and \sehgal\ respectively, and do not include the reionization kSZ component that we are most interested in here. The black solid line shows the prediction of the kSZ 4-point signal from the \citet{alvarez16} simulations. Thus we believe the measurement is dominated by foregrounds, as discussed in Sections~\ref{sec:fg_sims} and \ref{sec:results}, so should be interpreted only as an upper limit on the reionization signal. 
    }\label{fig:dr61}
\end{figure*}

\begin{figure*}
 \includegraphics[width=0.9\columnwidth]{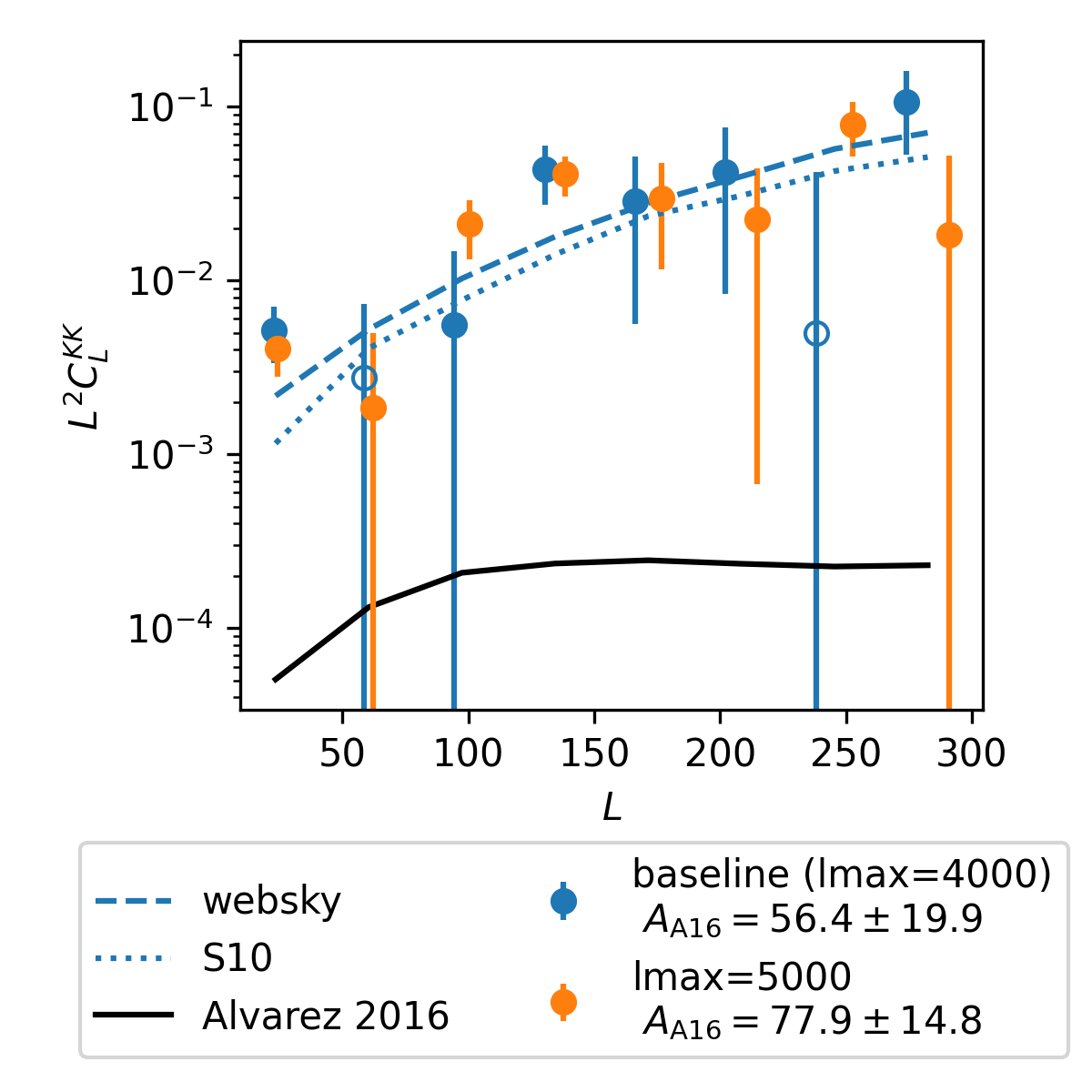}
 \includegraphics[width=0.9\columnwidth]{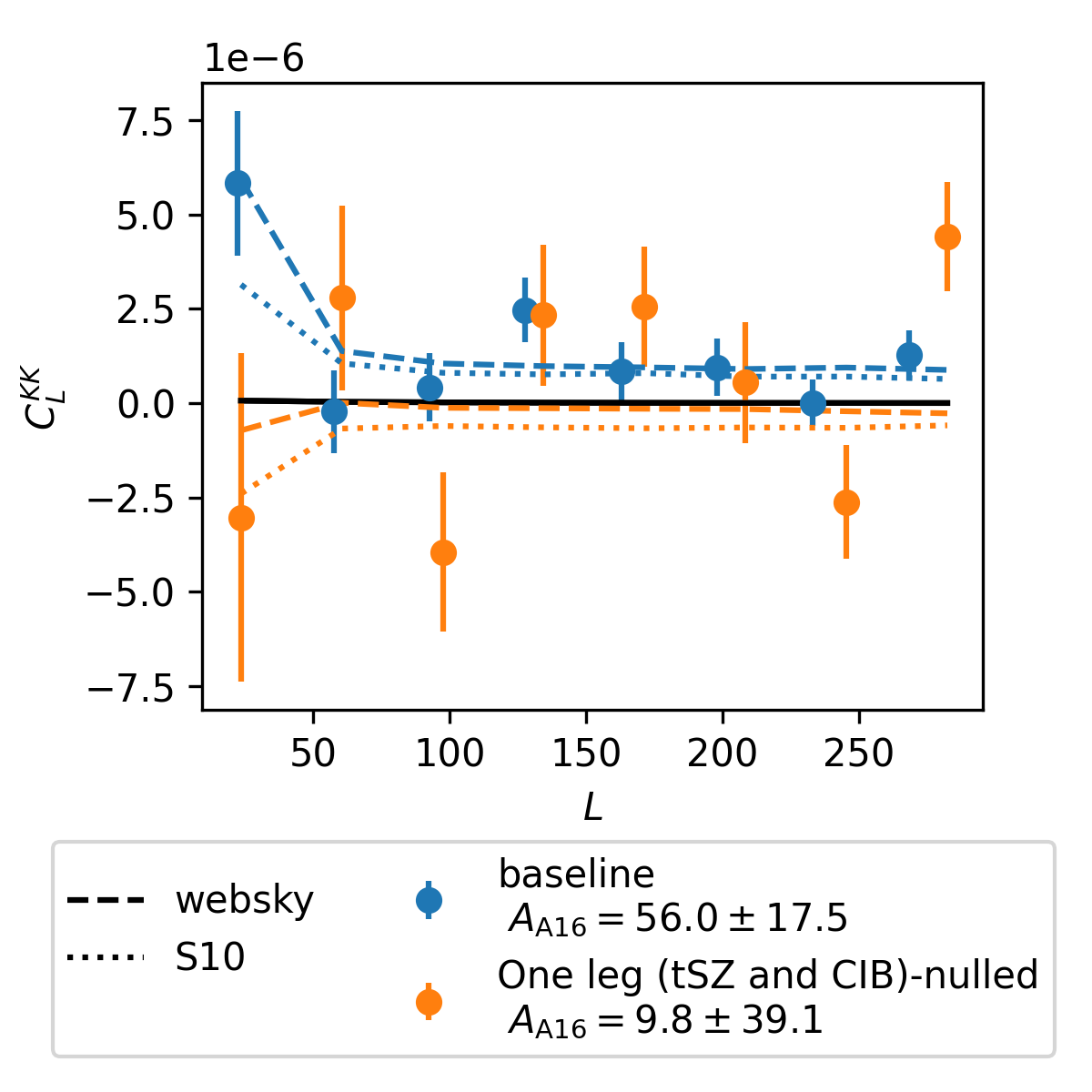}
 \caption{Measurements of $C_L^{KK}$ from ACT and \planck\ data (combined via a harmonic space ILC), highlighting alternative choices for analysis. Negative data points are indicated by open symbols. Left panel: The blue data points show our baseline measurement which uses CMB temperature $\lmax=4000$, while the orange data points use $\lmax=5000$. Right panel: The blue data points show our baseline measurement which  uses a minimum variance harmonic-space ILC in all trispectrum legs, while for the orange points, one of the four trispectrum legs has tSZ and CIB frequency spectra nulled (see \sect{sec:freqclean}). Note that we have switched to a linear scale here given the expectation that the tSZ and CIB-nulled measurement can be negative, and find that removing the $L^2$ scaling allows for a clearer view of the data points (and also results in a very slightly different amplitude fit for the baseline case). As elsewhere, the black solid line shows the prediction of the kSZ 4-point signal from the \citet{alvarez16} simulations and the dashed and solid lines show the corresponding predictions of the foreground bias from \websky\ and \sehgal\ respectively.} 
 \label{fig:dr62}
\end{figure*}
\footnotetext{this difference in $L$-scaling explains the small change in fitted amplitude for the baseline case.}

We show in \fig{fig:dr61} the measurement from ACT DR6 + \planck, combined via a harmonic space ILC, as described in \sect{sec:data}. 
We calculate the best-fit re-scaling amplitude of the \citet{alvarez16} theory prediction, which is reported in the legend.
Following \citet{smith17} and \citet{ferraro18}, we use $\lmin=3000$. For our baseline measurement we use $\lmax=4000$ 
. We find consistent results when we use $\lmax=5000$, but the reduction in uncertainty is modest (see left panel of \fig{fig:dr62}), and as shown in \sect{sec:fg_sims} the expected foreground biases are smaller for $\lmax=4000$. 

The measurement is generally well above the \citet{alvarez16} theory prediction, and is more consistent with the simulation predictions of the contamination from extra-galactic foregrounds (dashed and dotted lines from the \websky\ and \sehgal\ simulations respectively). The plotted errorbars are the diagonal of a covariance matrix calculated by running the $C_L^{KK}$ estimator on the simulations described in \sect{sec:bias_sims}. 

In the right panel of \fig{fig:dr62} we show cases where we use foreground-cleaned temperature in one of the four trispectrum legs. Orange points show the case where frequency spectra for tSZ and CIB are nulled in one of the trispectrum legs. The $C_L^{KK}$ uncertainty roughly doubles for this case, however, in our scenario, where the upper limit is  dominated by the foreground biases, this noise cost is not particularly relevant. The result from the baseline (ILC-only) case has the advantage that the foreground bias is very likely positive, hence the measurement can be straightforwardly interpreted as an upper limit. In contrast, the case where we deproject tSZ and CIB from one leg of the estimator is no longer an auto-correlation, so the residual foreground bias need not be positive. 

We do not perform a model fit due to the significant uncertainties in the foreground template, as can be inferred from the difference in the predictions from the \websky\ and \sehgal\ simulations. We discuss improvements to data and methodology designed to enable robust modeling of the signal in \sect{sec:discussion}. 

When we apply lensing-hardening, we obtain measurements consistent with the baseline case ($\Aksz = 77\pm33$). In the current regime where other foregrounds are dominant, lensing-hardening does not yet improve our upper limits. We find also that point-source hardening results in a very large increase in uncertainty (a factor of 20 with respect to the baseline case). We conclude that  our current data do not yet warrant the use of point-source hardening, and we do not present measurements using it here


\subsection{Robustness of $N^0$ and mean-field subtraction}\label{sec:N0}



\begin{figure*}
    \centering
    \includegraphics[width=0.45\textwidth]{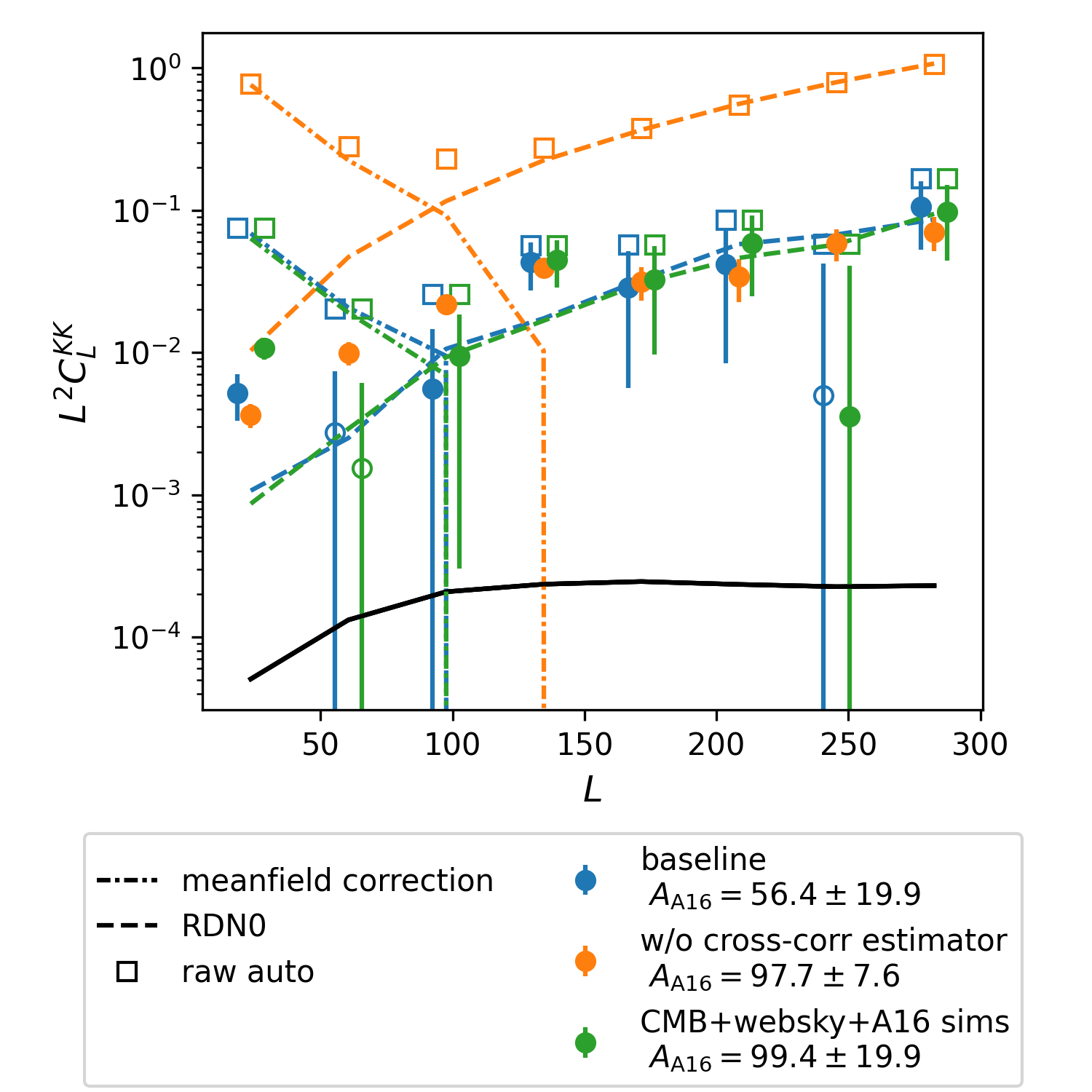}
    \includegraphics[width=0.45\textwidth]{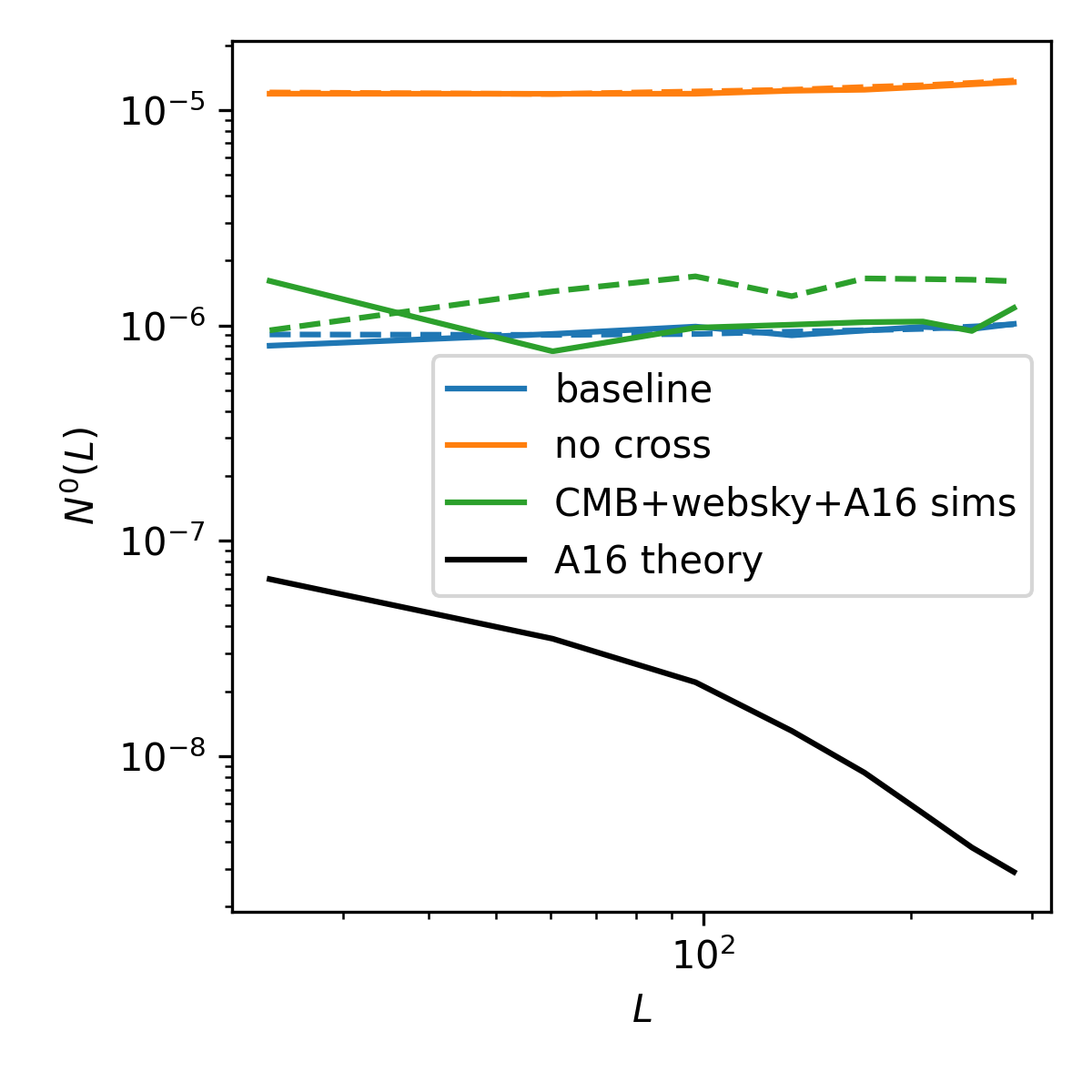}
    \caption{Demonstration of the importance of the cross-correlation based estimator of \citet{madhavacheril20}, and using simulations with the correct power spectrum, for our measurement. Left panel: The orange data points, which do not restrict to cross-correlations in the trispectrum estimation, result in a worse upper-limit compared to the baseline case (blue data points). Green data points do use the cross-correlation-based estimator, but do not use a set simulations that have power spectrum tuned to the data (see \sect{sec:N0} for details). Negative data points are indicated by open symbols. Right panel: $N^0$ estimates for the same three cases. RD$N^0$ is shown in solid lines, and MC$N^0$ in dashed lines. For the baseline case in blue, there is good agreement between  RD$N^0$ and MC$N^0$, arising from the good agreement in the power spectrum between simulations and data. For the cases where the cross-correlation estimator is not used the $N^0$ is much higher, making this estimator more susceptible to small errors in the $N^0$ correction. For the ``CMB+websky+A16'' case, we do see significant differences between MC$N^0$ and RD$N^0$.}
    \label{fig:dr63}
\end{figure*}

We discuss here the importance for our measurement of using the cross-correlation based estimator (as discussed in \sect{sec:n0andmf}), and a realization-dependent $N^0$ calculation. The left-hand panel of \fig{fig:dr63} demonstrates the improvement when using the cross-correlation based estimator. Both the mean-field and $N^0$ are much smaller, and given the lack of dependence on the observational noise, presumably more accurate, leading to a tighter upper limit when using the cross-correlation based estimator (blue data points). We also show a set of green points where we do use the cross-correlation-based estimator, but the simulations used for the mean-field and $N^0$ bias calculations are not matched to data. Instead they are generated assuming a power spectrum that is the sum of a primary CMB theory prediction, a foreground power spectrum contribution estimated from \websky, and a reionization kSZ power spectrum contribution estimated from the \citet{alvarez16} simulation (labelled ``CMB+websky+A16 sims''). We see a larger amplitude for both of these cases, which we consider to be less robust than our baseline analysis. 

In the right-hand panel of \fig{fig:dr63} we focus on the $N^0$ for these same three cases. The solid lines show RD$N^0$ and dashed lines MC$N^0$. For our baseline case, where the simulations are tuned to match the data power spectrum, we see good agreement between MC$N^0$ and RD$N^0$. For the case where the cross-correlation estimator is not used, we also see good agreement, but given the correction is much larger in this case, and two or three orders of magnitude above the signal, one would need to validate the correction extremely carefully. This motivates our choice to use the cross-correlation-only estimator in this work.  
Again we also show the ``CMB+websky+A16 sims'' case, which reveals differences between RD$N^0$ and MC$N^0$. If one used the naive MC$N^0$ to correct the signal in this case, one would significantly bias the $C_L^{KK}$ signal low. The use of RD$N^0$ in this case does lead to better agreement with the baseline case which we consider most robust. \



\section{Conclusions}\label{sec:discussion}

We  presented a limit on the kSZ trispectrum from wide-field CMB data which combines observations from ACT DR6 and \planck. Estimating this trispectrum from this dataset has presented various challenges, especially the levels of foreground and CMB lensing contamination, and uncertaintes in the mean-field and $N^0$ corrections that are potentially large compared to the signal of interest. Similarly to \citet{raghunathan24}, we find that the contamination due to foregrounds dominates over the reionization signal. 

We address these challenges by presenting a range of mitigation schemes. For lensing and extra-galactic foregrounds, we demonstrated the performance of bias-hardening for the first time in this context. Bias hardening against CMB lensing can very effectively remove the bias due to CMB lensing, and is likely to be useful beyond the kSZ trispectrum measurement, for example, in the `projected fields' method of kSZ measurement \citep{dore04, hill16, ferraro16}. 

We also proposed new trispectrum estimators where only a single leg has foreground frequency spectra nulled, which have significantly better performance than nulling all trispectrum legs. For the mean-field and $N^0$, also inspired by CMB lensing approaches, we demonstrate the utility of cross-correlation estimators and realization-dependent $N^0$, as well as the importance of having simulations whose power spectrum match the data. 

With these approaches in hand, we measure a trispectrum from ACT DR6 + \planck\ data that has an amplitude roughly 50 times the expected reionization signal from the \citet{alvarez16} simulations. This signal is roughly consistent with the expectation from extra-galactic foregrounds, as we have demonstrated using two independent foreground simulations (from \citealt{websky} and \citealt{sehgal10}). Thus we are not yet in a position to constrain realistic reionization scenarios. 

As this work was being finalized, \citet{raghunathan24} presented  an upper-limit on the reionization kSZ trispectrum from SPT+\textit{Herschel}-SPIRE data that is $\sim 5$ to $10$ times tighter than this work.  The authors argued that their inclusion of high resolution, high frequency data from \textit{Herschel}-SPIRE mitigated CIB contamination.  Even more importantly, since the SPT maps have roughly five times lower noise than the ACT maps used here (though over much smaller area), the authors were able to apply much stricter flux thresholds for point-sources and clusters which are the main contributors to foreground contamination.  In this paper we have emphasized methodologies for robust first detections of the small signatures of reionization on the CMB trispectrum in the future, and demonstrated them on the ACT DR6+\planck\ data.  We anticipate that techniques described here will be essential:   bias-hardening, RD$N^0$ and cross-correlation-based estimators.

We believe the constraint presented here can be improved by incorporating further ACT data (in particular data taken during daytime if we can demonstrate adequate systematics control), and pursuing more aggressive source and cluster finding. An analysis that more optimally takes into account the noise variations in the ACT data (both when performing the ILC, and in the filtering of the temperature modes entering the trispectrum estimator) will also be a focus of future work.
Beyond ACT, upcoming data from Simons Observatory and CMB-S4, which will have an order of magnitude lower noise, and additional frequency channels, have the potential to make the kSZ trispectrum a highly informative probe of reionization \citep{alvarez20}.
\appendix

\section{kSZ in the AMBER simulations}\label{app:amber}

\begin{figure*}
    \centering
    \includegraphics[width=0.95\textwidth]{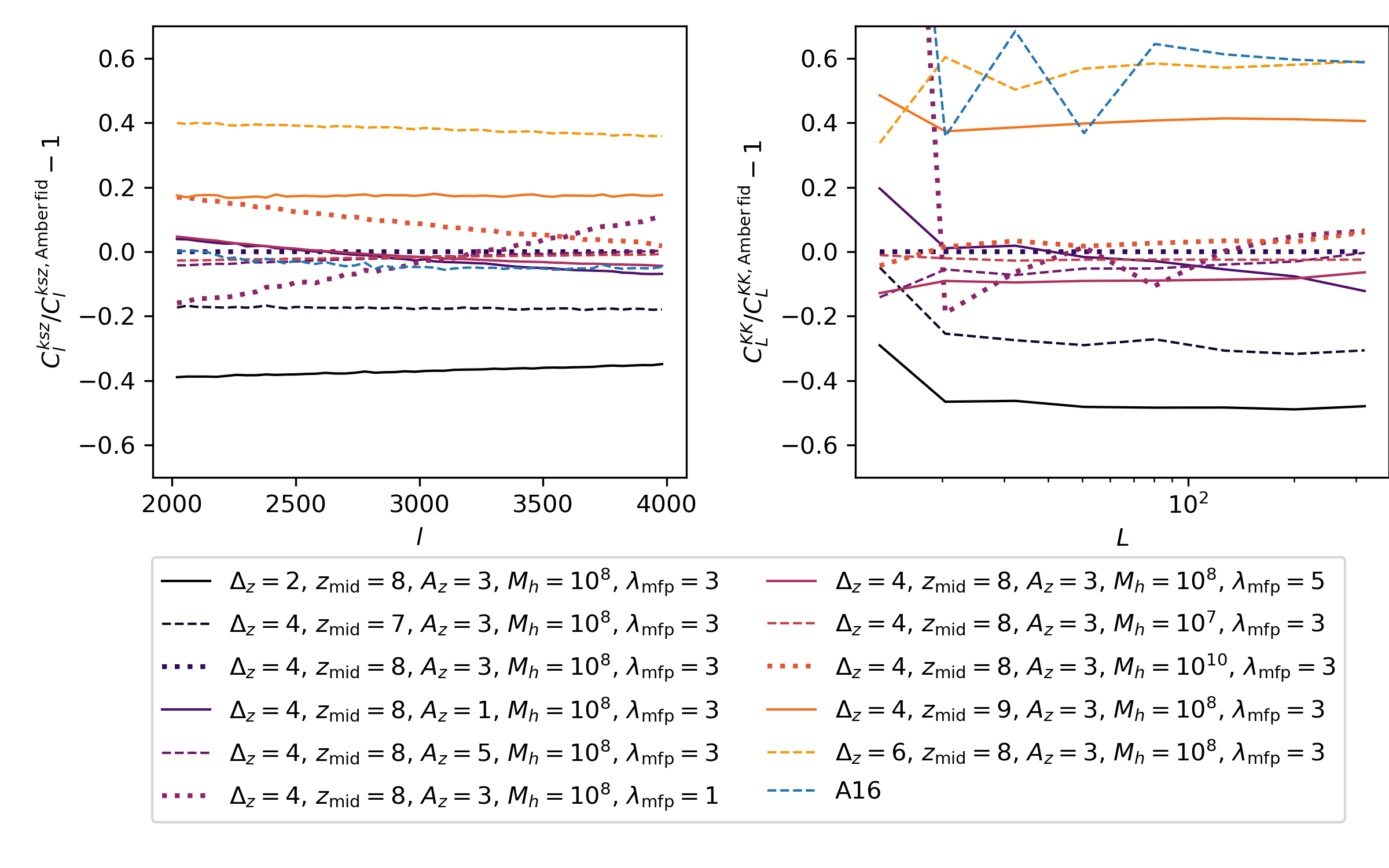}
    \caption{Predictions of the kSZ power spectrum (left panel), and trispectrum (right-panel) from reionization, using the AMBER simulations \citep{chen23,trac22}. We show the fractional difference with respect to a fiducial AMBER simulation (the third line in the legend). We also show the prediction based on the \citet{alvarez16} simulations (blue dashed lines). For the trispectrum, below $L=10$ the fractional differences become very noisy thus we only show the range $L>10$ for clarity.}
    \label{fig:amber_theory}
\end{figure*}

\fig{fig:amber_theory} shows the kSZ power spectrum (left-panel) and trispectrum (right-panel) measured from the AMBER simuations \citep{trac22,chen23}. In both cases the predictions are normalized by the measurement from the fiducial AMBER simulation\footnote{i.e. not the \citet{alvarez16} simulation, which we use as a fiducial template throughout the paper}. The AMBER simulations vary several reionization parameters:
\begin{itemize}
    \item $\Delta z$: The duration of reionization, defined as $z_{10}-z_{90}$, where $z_{10}$ and $z_{90}$ are the redshifts where the Universe is 10\% and 90\% ionized respectively.
    \item $\zmid$: The redshift where the Universe is 50\% ionized.
    \item $A_z$: An assymetry parameter defined $(z_{10}-\zmid)/(\zmid-z_{90})$.
    \item $M_h$: the minimum halo mass hosting ionizing sources (specified in \fig{fig:amber_theory} in $\mathrm{M}_*$).
    \item $\lambda_{\mathrm{mpf}}$: The mean free path of ionizing photons (specified in \fig{fig:amber_theory} in  $\mathrm{Mpc}/h$).
\end{itemize}
As argued in \sect{sec:theory}, we see a strong dependence for both power spectrum and trispectrum on $\Delta z$ and $\zmid$, with longer or earlier reionization increasing both signals. We show also, as the blue dashed line, the prediction from \citet{alvarez16} that we use as a template throughout the paper. This predicts a higher power spectrum and trispectrum than most of the AMBER variations which can be explained by the relatively high choice of $\zmid=10$. The other parameters varied in AMBER have less impact on the kSZ power spectrum and trispectrum, but see Figures 10 and 13 in \citet{chen23} for a more detailed view.

We see also that the response to changing these parameters is different for the power spectrum and trispectrum, demonstrating the complementary information in the two statistics (see \citealt{alvarez20} for forecasts of degeneracy-breaking power of these two statistics). 

When foregrounds are taken into account, which also affect the two signals differently, the gain from including the trispectrum information can be even greater. While the same is true in theory for the other varied parameters, $A_z$, $M_h$ and $\lambda_{\mathrm{mpf}}$, the responses for the trispectrum are relatively small, and thus will be difficult to detect with CMB data. 

\section{CMB temperature  \lowercase{$l$}-dependence of foreground biases}\label{sec:fgmax}

\begin{figure*}
    \centering
    \includegraphics[width=0.98\textwidth]{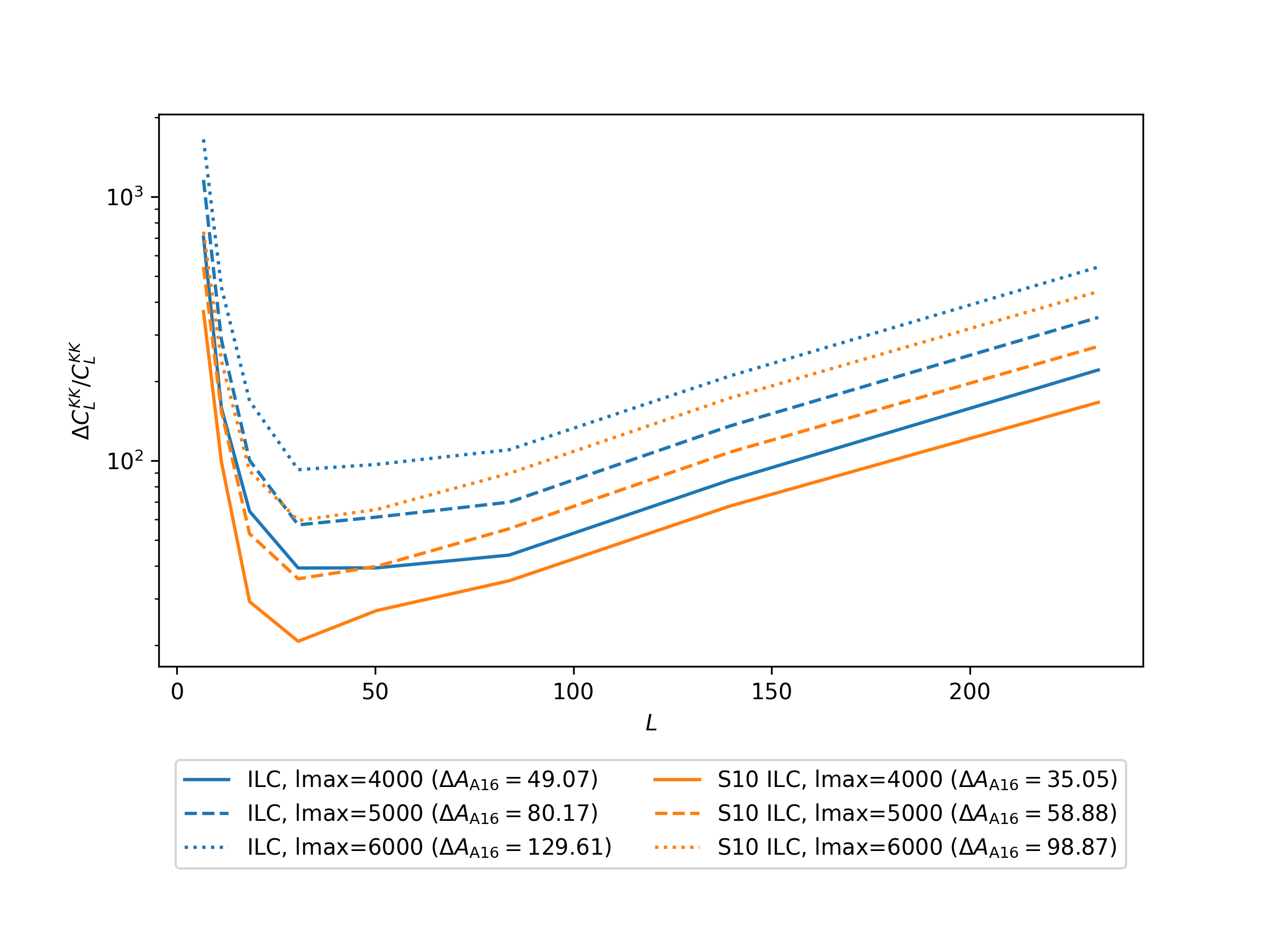}
    \caption{Expected fractional foreground biases from the \websky\ (blue lines) and \sehgal\ (orange lines) simulations. Here we show the impact on the foreground bias of changing the range of scales $l$ entering the quadratic estimator. }
    \label{fig:fg2}
\end{figure*}

In \fig{fig:fg2} we demonstrate the dependence of the fractional foreground bias, on the maximum CMB temperature multipole, $l$ used, for the ILC case. For both \websky\ and \sehgal, there is a significant increase in the foreground biases when increasing $\lmax$, hence we use $\lmax=4000$ for our baseline analysis.

\section{Normalization conventions}\label{app:norm}

While we have motivated the quadratic estimator for $K$ by analogy with CMB lensing, the choice of normalisation of the $K$ estimator is less obvious than in the case of CMB lensing, where one is reconstructing the lensing potential $\phi$. In the main body of this work, we have adopted a convention that is closely related to the \citet{sailer20} source estimator, mainly for convenience when adapting the CMB lensing codes to measure and de-bias the $K$ statistic (and include lensing-hardening etc.). The \citet{sailer20} source estimator is defined:
\begin{equation}
s^{\mathrm{S20}}_{\vecL} = N^s(L) \int \frac{\dxsq{\vecl}}{(2\pi)^2} 
\frac{u_l u_{|\vecL-\vecl|} T_{\vecl} T_{\vecL-\vecl} }{2 u_L C_l^{\mathrm{tot}}C_{|\vecL-\vecl|}^{\mathrm{tot}} }
\end{equation}
where $u_L$ is the source profile and
\begin{equation}
\left[N^s(L)\right]^{-1} = \int \frac{\dxsq{\vecl}}{(2\pi)^2} \frac{1}{2 C_l^{\mathrm{tot}} C_{|\vecL-\vecl|}^{\mathrm{tot}}} \left[\frac{u_l u_{|\vecL-\vecl|}}{u_L}\right]^2.
\end{equation}
After substituting $u_L=\sqrt{C_l^{\mathrm{pkSZ}}}$, our $K$ estimator is similar, except we divide out dependence on $u_L$, since our measurement should not depend on $C_l^{\mathrm{pkSZ}}$ at large scales:
\begin{align}
    K_{\vecL} &= \left[ \int \frac{\dxsq{\vecl}}{(2\pi)^2} u_l u_{L-l} W_l W_{|\vecL-\vecl|}  \right]^{-1} \\
    &\times \int \frac{\dxsq{\vecl}}{(2\pi)^2} W_l W_{|\vecL-\vecl|} T_{\vecl}T_{\vecL-\vecl}.
\end{align}
The normalisation factor here is somewhat different to that chosen by \citet{smith17}, who include a $C_l^{\mathrm{tot}}$ factor in their normalisation, aiming to construct a $K$ statistic that is a measure of the kSZ non-Gaussianity relative to the total power spectrum:
\begin{equation}
    \bar{K} = \int \frac{\dxsq{\vecl}}{(2\pi)^2} W_s^2(l)C_l^{\mathrm{tot}}.
\end{equation}
$C_l^{\mathrm{tot}}$ is however both survey-dependent (if it includes noise) and the signal component is quite uncertain in the $l>3000$ regime we are considering. 
This would motivate a normalisation where one simply divides out the effect of the filtering, which we call $K_{\mathrm{filter}}$, defined
\begin{align}
    K_{\mathrm{filter}} &= \int \frac{\dx{^2l}}{(2\pi)^2} (W_l^{\mathrm{eff}}W_{|L-l|}^{\mathrm{eff}})\label{eq:K_filter}\\
    &\approx \int \frac{\dx{^2l}}{(2\pi)^2} (W_l^{\mathrm{eff}})^2\\
    &\approx \int \frac{l\dx{l}}{2\pi} (W_l^{\mathrm{eff}})^2.
\end{align}
In the second line we have assumed the squeezed limit $L<<l$.

\citet{raghunathan24} adopt a different normalisation again:
\begin{equation}
    K^{R24}_L = \bar{F}^{-1} \int \frac{\dxsq {\vecl}}{(2\pi)^2} W_l W_{|\vecL-\vecl|} T_{\vecl}T_{\vecL-\vecl}
\end{equation}
where 
\begin{equation}
    \bar{F} = \int W_l^2 B^2_{l,\mathrm{eff}} F_l \dx{\ln{l}}.
\end{equation}
Note for simplicity we have neglected their $f_{\mathrm{sky}}$ corrections, as we have in the discussion above. $B^2_{l,\mathrm{eff}}$ denotes their effective beam, and can be considered as part of an effective filter $W_l^{\mathrm{eff}}$. 

\begin{figure*}
    \centering
    \includegraphics[width=0.45\textwidth]{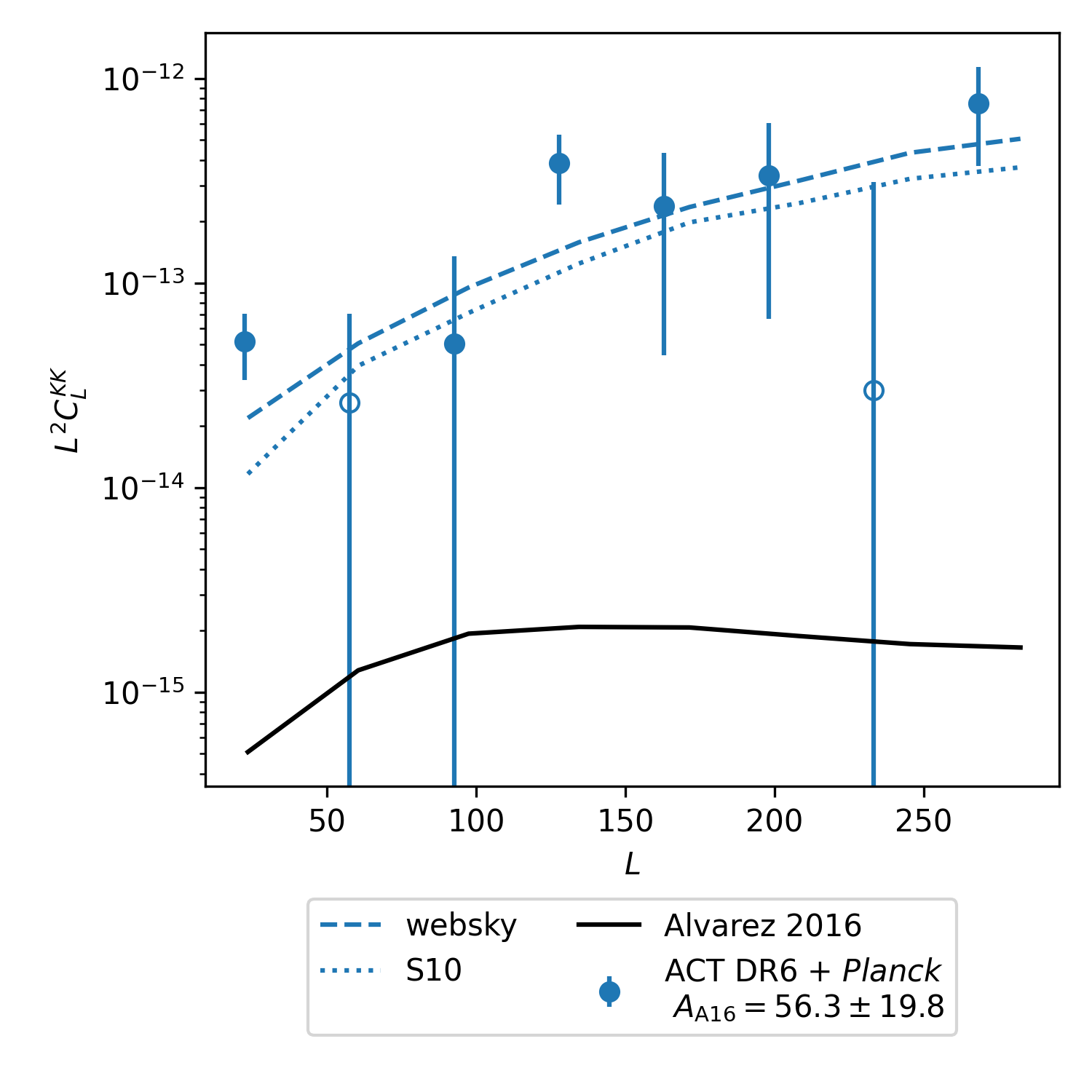}
    \includegraphics[width=0.45\textwidth]{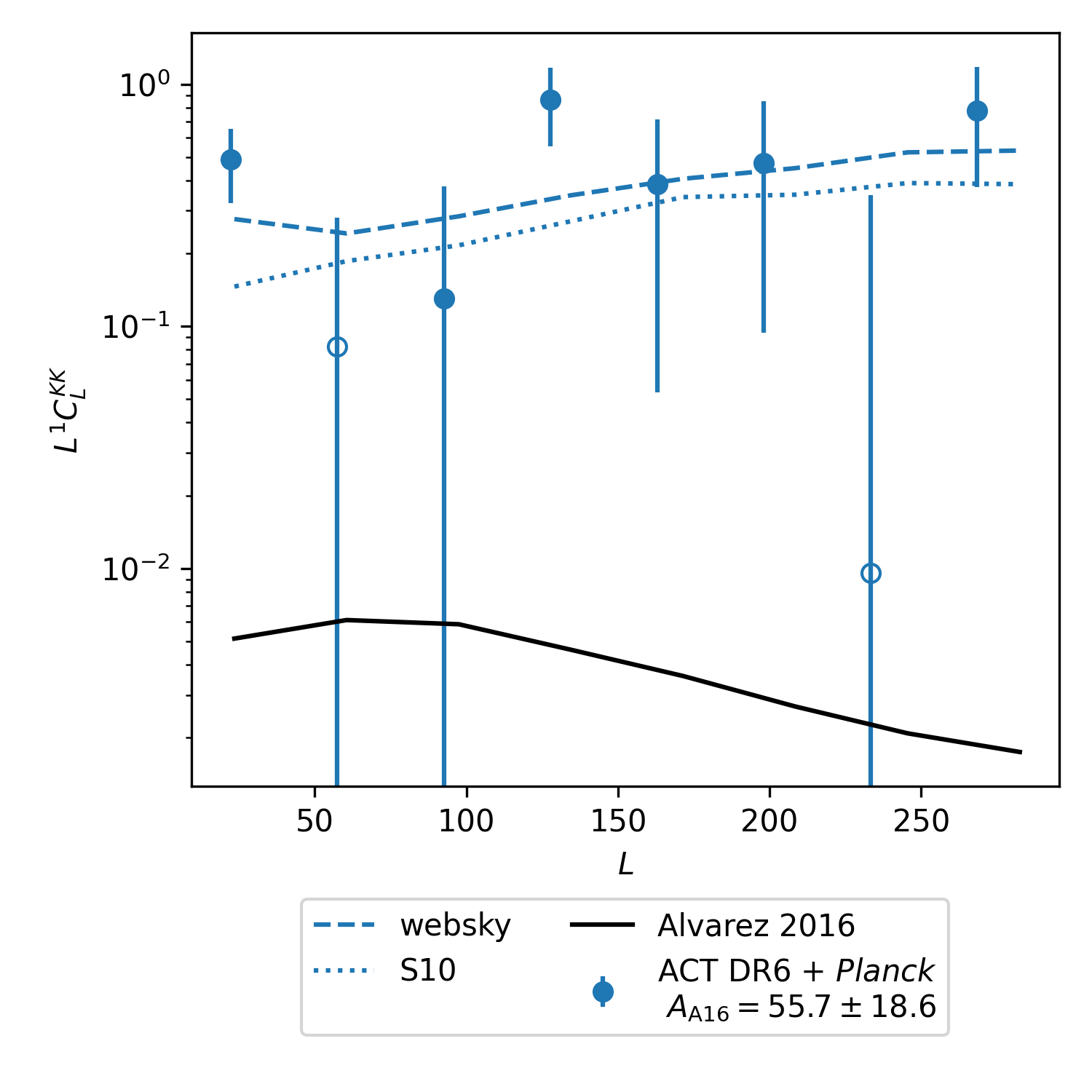}
    \caption{Two versions of this work's \fig{fig:dr61} with alternative normalisations, in case useful for comparison purpuses. In the left panel, we normalise $K$ with $K_{\mathrm{filter}}$ (defined in \eqn{eq:K_filter}. In the right hand panel, we use the \citet{raghunathan24} normalisation, and also their $L$-scaling of the y-axis.}
    \label{fig:dr6_norm}
\end{figure*}

In case useful for comparison purposes, we show in \fig{fig:dr6_norm} versions of our \fig{fig:dr61} using the $K_{\mathrm{filter}}$ and $\bar{F}$ normalisations.

\section*{Acknowledgements}
Support for ACT was through the U.S.~National Science Foundation through awards AST-0408698, AST-0965625, and AST-1440226 for the ACT project, as well as awards PHY-0355328, PHY-0855887 and PHY-1214379. Funding was also provided by Princeton University, the University of Pennsylvania, and a Canada Foundation for Innovation (CFI) award to UBC. ACT operated in the Parque Astron\'omico Atacama in northern Chile under the auspices of the Agencia Nacional de Investigaci\'on y Desarrollo (ANID). The development of multichroic detectors and lenses was supported by NASA grants NNX13AE56G and NNX14AB58G. Detector research at NIST was supported by the NIST Innovations in Measurement Science program. Computing for ACT was performed using the Princeton Research Computing resources at Princeton University, the National Energy Research Scientific Computing Center (NERSC), and the Niagara supercomputer at the SciNet HPC Consortium. SciNet is funded by the CFI under the auspices of Compute Canada, the Government of Ontario, the Ontario Research Fund–Research Excellence, and the University of Toronto. We thank the Republic of Chile for hosting ACT in the northern Atacama, and the local indigenous Licanantay communities whom we follow in observing and learning from the night sky.

CS acknowledges support from the Agencia Nacional de Investigaci\'on y Desarrollo (ANID) through Basal project FB210003. EC acknowledges support from the European Research Council (ERC) under the European Union’s Horizon 2020 research and innovation programme (Grant agreement No. 849169). OD acknowledges support from a SNSF Eccellenza Professorial Fellowship (No. 186879). JCH acknowledges support from NSF grant AST-2108536, DOE grant DE-SC0011941, NASA grants 21-ATP21-0129 and 22-ADAP22-0145, the Sloan Foundation, and the Simons Foundation. R.~H. acknowledges support from CIFAR, the Azrieli and Alfred. P. Sloan foundations, and the NSERC Discovery Grants program. NM acknowledges the support of the Royal Society, grant URFR1221806.

\section*{Data Availability}

We will make the baseline analysis data-points and covariance matrices available upon publication of this work.



\bibliographystyle{mnras}
\bibliography{refs} 

\begin{thebibliography}{}
\makeatletter
\relax
\def\mn@urlcharsother{\let\do\@makeother \do\$\do\&\do\#\do\^\do\_\do\%\do\~}
\def\mn@doi{\begingroup\mn@urlcharsother \@ifnextchar [ {\mn@doi@}
  {\mn@doi@[]}}
\def\mn@doi@[#1]#2{\def\@tempa{#1}\ifx\@tempa\@empty \href
  {http://dx.doi.org/#2} {doi:#2}\else \href {http://dx.doi.org/#2} {#1}\fi
  \endgroup}
\def\mn@eprint#1#2{\mn@eprint@#1:#2::\@nil}
\def\mn@eprint@arXiv#1{\href {http://arxiv.org/abs/#1} {{\tt arXiv:#1}}}
\def\mn@eprint@dblp#1{\href {http://dblp.uni-trier.de/rec/bibtex/#1.xml}
  {dblp:#1}}
\def\mn@eprint@#1:#2:#3:#4\@nil{\def\@tempa {#1}\def\@tempb {#2}\def\@tempc
  {#3}\ifx \@tempc \@empty \let \@tempc \@tempb \let \@tempb \@tempa \fi \ifx
  \@tempb \@empty \def\@tempb {arXiv}\fi \@ifundefined
  {mn@eprint@\@tempb}{\@tempb:\@tempc}{\expandafter \expandafter \csname
  mn@eprint@\@tempb\endcsname \expandafter{\@tempc}}}

\bibitem[\protect\citeauthoryear{{Ade} et~al.,}{{Ade} et~al.}{2019}]{sogoals}
{Ade} P.,  et~al., 2019, \mn@doi [\jcap] {10.1088/1475-7516/2019/02/056}, \href
  {https://ui.adsabs.harvard.edu/abs/2019JCAP...02..056A} {2019, 056}

\bibitem[\protect\citeauthoryear{Alvarez}{Alvarez}{2016}]{alvarez16}
Alvarez M.~A.,  2016, \mn@doi [The Astrophysical Journal]
  {10.3847/0004-637x/824/2/118}, 824, 118

\bibitem[\protect\citeauthoryear{{Alvarez}, {Ferraro}, {Hill}, {Hlo{\v{z}}ek}
  \& {Ikape}}{{Alvarez} et~al.}{2021}]{alvarez20}
{Alvarez} M.~A.,  {Ferraro} S.,  {Hill} J.~C.,  {Hlo{\v{z}}ek} R.,   {Ikape}
  M.,  2021, \mn@doi [\prd] {10.1103/PhysRevD.103.063518}, \href
  {https://ui.adsabs.harvard.edu/abs/2021PhRvD.103f3518A} {103, 063518}

\bibitem[\protect\citeauthoryear{{Atkins} et~al.,}{{Atkins}
  et~al.}{2023}]{atkins23}
{Atkins} Z.,  et~al., 2023, \mn@doi [\jcap] {10.1088/1475-7516/2023/11/073},
  \href {https://ui.adsabs.harvard.edu/abs/2023JCAP...11..073A} {2023, 073}

\bibitem[\protect\citeauthoryear{{Battaglia}, {Natarajan}, {Trac}, {Cen}  \&
  {Loeb}}{{Battaglia} et~al.}{2013}]{battaglia13}
{Battaglia} N.,  {Natarajan} A.,  {Trac} H.,  {Cen} R.,   {Loeb} A.,  2013,
  \mn@doi [\apj] {10.1088/0004-637X/776/2/83}, \href
  {https://ui.adsabs.harvard.edu/abs/2013ApJ...776...83B} {776, 83}

\bibitem[\protect\citeauthoryear{{Calabrese} et~al.,}{{Calabrese}
  et~al.}{2014}]{calabrese14}
{Calabrese} E.,  et~al., 2014, \mn@doi [\jcap] {10.1088/1475-7516/2014/08/010},
  \href {https://ui.adsabs.harvard.edu/abs/2014JCAP...08..010C} {2014, 010}

\bibitem[\protect\citeauthoryear{{Chen}, {Trac}, {Mukherjee}  \& {Cen}}{{Chen}
  et~al.}{2023}]{chen23}
{Chen} N.,  {Trac} H.,  {Mukherjee} S.,   {Cen} R.,  2023, \mn@doi [\apj]
  {10.3847/1538-4357/ac8481}, \href
  {https://ui.adsabs.harvard.edu/abs/2023ApJ...943..138C} {943, 138}

\bibitem[\protect\citeauthoryear{{Chluba}, {Hill}  \& {Abitbol}}{{Chluba}
  et~al.}{2017}]{chluba17}
{Chluba} J.,  {Hill} J.~C.,   {Abitbol} M.~H.,  2017, \mn@doi [\mnras]
  {10.1093/mnras/stx1982}, \href
  {https://ui.adsabs.harvard.edu/abs/2017MNRAS.472.1195C} {472, 1195}

\bibitem[\protect\citeauthoryear{Choi et~al.,}{Choi et~al.}{2020}]{choi20}
Choi S.~K.,  et~al., 2020, \mn@doi [Journal of Cosmology and Astroparticle
  Physics] {10.1088/1475-7516/2020/12/045}, 2020, 045–045

\bibitem[\protect\citeauthoryear{{Coulton} et~al.,}{{Coulton}
  et~al.}{2024}]{coulton24}
{Coulton} W.,  et~al., 2024, \mn@doi [\prd] {10.1103/PhysRevD.109.063530},
  \href {https://ui.adsabs.harvard.edu/abs/2024PhRvD.109f3530C} {109, 063530}

\bibitem[\protect\citeauthoryear{{Darwish}, {Sherwin}, {Sailer}, {Schaan}  \&
  {Ferraro}}{{Darwish} et~al.}{2021}]{darwish21}
{Darwish} O.,  {Sherwin} B.~D.,  {Sailer} N.,  {Schaan} E.,   {Ferraro} S.,
  2021, \mn@doi [arXiv e-prints] {10.48550/arXiv.2111.00462}, \href
  {https://ui.adsabs.harvard.edu/abs/2021arXiv211100462D} {p. arXiv:2111.00462}

\bibitem[\protect\citeauthoryear{Dore, Hennawi  \& Spergel}{Dore
  et~al.}{2004}]{dore04}
Dore O.,  Hennawi J.~F.,   Spergel D.~N.,  2004, \mn@doi [The Astrophysical
  Journal] {10.1086/382946}, 606, 46–57

\bibitem[\protect\citeauthoryear{{Eisenstein} et~al.,}{{Eisenstein}
  et~al.}{2023}]{eisenstein23}
{Eisenstein} D.~J.,  et~al., 2023, \mn@doi [arXiv e-prints]
  {10.48550/arXiv.2306.02465}, \href
  {https://ui.adsabs.harvard.edu/abs/2023arXiv230602465E} {p. arXiv:2306.02465}

\bibitem[\protect\citeauthoryear{{Ferraro} \& {Smith}}{{Ferraro} \&
  {Smith}}{2018}]{ferraro18}
{Ferraro} S.,  {Smith} K.~M.,  2018, \mn@doi [\prd]
  {10.1103/PhysRevD.98.123519}, \href
  {https://ui.adsabs.harvard.edu/abs/2018PhRvD..98l3519F} {98, 123519}

\bibitem[\protect\citeauthoryear{{Ferraro}, {Hill}, {Battaglia}, {Liu}  \&
  {Spergel}}{{Ferraro} et~al.}{2016}]{ferraro16}
{Ferraro} S.,  {Hill} J.~C.,  {Battaglia} N.,  {Liu} J.,   {Spergel} D.~N.,
  2016, \mn@doi [\prd] {10.1103/PhysRevD.94.123526}, \href
  {https://ui.adsabs.harvard.edu/abs/2016PhRvD..94l3526F} {94, 123526}

\bibitem[\protect\citeauthoryear{{Finkelstein} et~al.,}{{Finkelstein}
  et~al.}{2023a}]{finkelstein23b}
{Finkelstein} S.~L.,  et~al., 2023a, \mn@doi [arXiv e-prints]
  {10.48550/arXiv.2311.04279}, \href
  {https://ui.adsabs.harvard.edu/abs/2023arXiv231104279F} {p. arXiv:2311.04279}

\bibitem[\protect\citeauthoryear{{Finkelstein} et~al.,}{{Finkelstein}
  et~al.}{2023b}]{finkelstein23a}
{Finkelstein} S.~L.,  et~al., 2023b, \mn@doi [\apjl]
  {10.3847/2041-8213/acade4}, \href
  {https://ui.adsabs.harvard.edu/abs/2023ApJ...946L..13F} {946, L13}

\bibitem[\protect\citeauthoryear{Gorce, Douspis  \& Salvati}{Gorce
  et~al.}{2022}]{gorce22}
Gorce A.,  Douspis M.,   Salvati L.,  2022, \mn@doi [Astronomy &amp;
  Astrophysics] {10.1051/0004-6361/202243351}, 662, A122

\bibitem[\protect\citeauthoryear{{Hanson}, {Challinor}, {Efstathiou}  \&
  {Bielewicz}}{{Hanson} et~al.}{2011}]{hanson11}
{Hanson} D.,  {Challinor} A.,  {Efstathiou} G.,   {Bielewicz} P.,  2011,
  \mn@doi [\prd] {10.1103/PhysRevD.83.043005}, \href
  {https://ui.adsabs.harvard.edu/abs/2011PhRvD..83d3005H} {83, 043005}

\bibitem[\protect\citeauthoryear{{Hill}, {Ferraro}, {Battaglia}, {Liu}  \&
  {Spergel}}{{Hill} et~al.}{2016}]{hill16}
{Hill} J.~C.,  {Ferraro} S.,  {Battaglia} N.,  {Liu} J.,   {Spergel} D.~N.,
  2016, \mn@doi [\prl] {10.1103/PhysRevLett.117.051301}, \href
  {https://ui.adsabs.harvard.edu/abs/2016PhRvL.117e1301H} {117, 051301}

\bibitem[\protect\citeauthoryear{Kesden, Cooray  \& Kamionkowski}{Kesden
  et~al.}{2003}]{kesden03}
Kesden M.,  Cooray A.,   Kamionkowski M.,  2003, \mn@doi [Phys. Rev. D]
  {10.1103/PhysRevD.67.123507}, 67, 123507

\bibitem[\protect\citeauthoryear{{Kogut}}{{Kogut}}{2003}]{kogut03}
{Kogut} A.,  2003, \mn@doi [\nar] {10.1016/j.newar.2003.09.029}, \href
  {https://ui.adsabs.harvard.edu/abs/2003NewAR..47..977K} {47, 977}

\bibitem[\protect\citeauthoryear{{Kusiak}, {Bolliet}, {Ferraro}, {Hill}  \&
  {Krolewski}}{{Kusiak} et~al.}{2021}]{kusiak21}
{Kusiak} A.,  {Bolliet} B.,  {Ferraro} S.,  {Hill} J.~C.,   {Krolewski} A.,
  2021, \mn@doi [\prd] {10.1103/PhysRevD.104.043518}, \href
  {https://ui.adsabs.harvard.edu/abs/2021PhRvD.104d3518K} {104, 043518}

\bibitem[\protect\citeauthoryear{{MacCrann} et~al.,}{{MacCrann}
  et~al.}{2023}]{maccrann23}
{MacCrann} N.,  et~al., 2023, \mn@doi [arXiv e-prints]
  {10.48550/arXiv.2304.05196}, \href
  {https://ui.adsabs.harvard.edu/abs/2023arXiv230405196M} {p. arXiv:2304.05196}

\bibitem[\protect\citeauthoryear{Madhavacheril \& Hill}{Madhavacheril \&
  Hill}{2018}]{madhavacheril18}
Madhavacheril M.~S.,  Hill J.~C.,  2018, \mn@doi [Physical Review D]
  {10.1103/physrevd.98.023534}, 98

\bibitem[\protect\citeauthoryear{{Madhavacheril}, {Smith}, {Sherwin}  \&
  {Naess}}{{Madhavacheril} et~al.}{2020}]{madhavacheril20}
{Madhavacheril} M.~S.,  {Smith} K.~M.,  {Sherwin} B.~D.,   {Naess} S.,  2020,
  \mn@doi [arXiv e-prints] {10.48550/arXiv.2011.02475}, \href
  {https://ui.adsabs.harvard.edu/abs/2020arXiv201102475M} {p. arXiv:2011.02475}

\bibitem[\protect\citeauthoryear{{McQuinn}, {Furlanetto}, {Hernquist}, {Zahn}
  \& {Zaldarriaga}}{{McQuinn} et~al.}{2005}]{mcquinn05}
{McQuinn} M.,  {Furlanetto} S.~R.,  {Hernquist} L.,  {Zahn} O.,   {Zaldarriaga}
  M.,  2005, \mn@doi [\apj] {10.1086/432049}, \href
  {https://ui.adsabs.harvard.edu/abs/2005ApJ...630..643M} {630, 643}

\bibitem[\protect\citeauthoryear{{Namikawa}, {Hanson}  \&
  {Takahashi}}{{Namikawa} et~al.}{2013}]{namikawa13}
{Namikawa} T.,  {Hanson} D.,   {Takahashi} R.,  2013, \mn@doi [\mnras]
  {10.1093/mnras/stt195}, \href
  {https://ui.adsabs.harvard.edu/abs/2013MNRAS.431..609N} {431, 609}

\bibitem[\protect\citeauthoryear{{Osborne}, {Hanson}  \& {Dor{\'e}}}{{Osborne}
  et~al.}{2014}]{osborne14}
{Osborne} S.~J.,  {Hanson} D.,   {Dor{\'e}} O.,  2014, \mn@doi [\jcap]
  {10.1088/1475-7516/2014/03/024}, \href
  {https://ui.adsabs.harvard.edu/abs/2014JCAP...03..024O} {2014, 024}

\bibitem[\protect\citeauthoryear{{Planck Collaboration} et~al.,}{{Planck
  Collaboration} et~al.}{2016}]{plankreion}
{Planck Collaboration} et~al., 2016, \mn@doi [\aap]
  {10.1051/0004-6361/201628897}, \href
  {https://ui.adsabs.harvard.edu/abs/2016A&A...596A.108P} {596, A108}

\bibitem[\protect\citeauthoryear{{Planck Collaboration} et~al.,}{{Planck
  Collaboration} et~al.}{2020}]{plancknpipe}
{Planck Collaboration} et~al., 2020, \mn@doi [\aap]
  {10.1051/0004-6361/202038073}, \href
  {https://ui.adsabs.harvard.edu/abs/2020A&A...643A..42P} {643, A42}

\bibitem[\protect\citeauthoryear{Qu et~al.}{Qu et~al.}{2024}]{qu23}
Qu F.~J.,  et~al., 2024, \mn@doi [Astrophys. J.] {10.3847/1538-4357/acfe06},
  962, 112

\bibitem[\protect\citeauthoryear{{Raghunathan} et~al.,}{{Raghunathan}
  et~al.}{2024}]{raghunathan24}
{Raghunathan} S.,  et~al., 2024, \mn@doi [arXiv e-prints]
  {10.48550/arXiv.2403.02337}, \href
  {https://ui.adsabs.harvard.edu/abs/2024arXiv240302337R} {p. arXiv:2403.02337}

\bibitem[\protect\citeauthoryear{{Reichardt} et~al.,}{{Reichardt}
  et~al.}{2021}]{reichardt21}
{Reichardt} C.~L.,  et~al., 2021, \mn@doi [\apj] {10.3847/1538-4357/abd407},
  \href {https://ui.adsabs.harvard.edu/abs/2021ApJ...908..199R} {908, 199}

\bibitem[\protect\citeauthoryear{{Remazeilles}, {Delabrouille}  \&
  {Cardoso}}{{Remazeilles} et~al.}{2011}]{remazeilles11}
{Remazeilles} M.,  {Delabrouille} J.,   {Cardoso} J.-F.,  2011, \mn@doi
  [\mnras] {10.1111/j.1365-2966.2010.17624.x}, \href
  {https://ui.adsabs.harvard.edu/abs/2011MNRAS.410.2481R} {410, 2481}

\bibitem[\protect\citeauthoryear{{Rotti} \& {Chluba}}{{Rotti} \&
  {Chluba}}{2021}]{rotti21}
{Rotti} A.,  {Chluba} J.,  2021, \mn@doi [\mnras] {10.1093/mnras/staa3292},
  \href {https://ui.adsabs.harvard.edu/abs/2021MNRAS.500..976R} {500, 976}

\bibitem[\protect\citeauthoryear{{Sailer}, {Schaan}  \& {Ferraro}}{{Sailer}
  et~al.}{2020}]{sailer20}
{Sailer} N.,  {Schaan} E.,   {Ferraro} S.,  2020, \mn@doi [\prd]
  {10.1103/PhysRevD.102.063517}, \href
  {https://ui.adsabs.harvard.edu/abs/2020PhRvD.102f3517S} {102, 063517}

\bibitem[\protect\citeauthoryear{{Santos}, {Cooray}, {Haiman}, {Knox}  \&
  {Ma}}{{Santos} et~al.}{2003}]{santos03}
{Santos} M.~G.,  {Cooray} A.,  {Haiman} Z.,  {Knox} L.,   {Ma} C.-P.,  2003,
  \mn@doi [\apj] {10.1086/378772}, \href
  {https://ui.adsabs.harvard.edu/abs/2003ApJ...598..756S} {598, 756}

\bibitem[\protect\citeauthoryear{{Savitzky} \& {Golay}}{{Savitzky} \&
  {Golay}}{1964}]{Savitzky64}
{Savitzky} A.,  {Golay} M.~J.~E.,  1964, \mn@doi [Analytical Chemistry]
  {10.1021/ac60214a047}, \href
  {https://ui.adsabs.harvard.edu/abs/1964AnaCh..36.1627S} {36, 1627}

\bibitem[\protect\citeauthoryear{{Sehgal}, {Bode}, {Das},
  {Hernandez-Monteagudo}, {Huffenberger}, {Lin}, {Ostriker}  \&
  {Trac}}{{Sehgal} et~al.}{2010}]{sehgal10}
{Sehgal} N.,  {Bode} P.,  {Das} S.,  {Hernandez-Monteagudo} C.,  {Huffenberger}
  K.,  {Lin} Y.-T.,  {Ostriker} J.~P.,   {Trac} H.,  2010, \mn@doi [\apj]
  {10.1088/0004-637X/709/2/920}, \href
  {https://ui.adsabs.harvard.edu/abs/2010ApJ...709..920S} {709, 920}

\bibitem[\protect\citeauthoryear{{Shaw}, {Rudd}  \& {Nagai}}{{Shaw}
  et~al.}{2012}]{shaw12}
{Shaw} L.~D.,  {Rudd} D.~H.,   {Nagai} D.,  2012, \mn@doi [\apj]
  {10.1088/0004-637X/756/1/15}, \href
  {https://ui.adsabs.harvard.edu/abs/2012ApJ...756...15S} {756, 15}

\bibitem[\protect\citeauthoryear{Smith \& Ferraro}{Smith \&
  Ferraro}{2017}]{smith17}
Smith K.~M.,  Ferraro S.,  2017, \mn@doi [Physical Review Letters]
  {10.1103/physrevlett.119.021301}, 119

\bibitem[\protect\citeauthoryear{Stein, Alvarez, Bond, Engelen  \&
  Battaglia}{Stein et~al.}{2020}]{websky}
Stein G.,  Alvarez M.~A.,  Bond J.~R.,  Engelen A.~v.,   Battaglia N.,  2020,
  \mn@doi [Journal of Cosmology and Astroparticle Physics]
  {10.1088/1475-7516/2020/10/012}, 2020, 012–012

\bibitem[\protect\citeauthoryear{{Thornton} et~al.,}{{Thornton}
  et~al.}{2016}]{thornton16}
{Thornton} R.~J.,  et~al., 2016, \mn@doi [\apjs] {10.3847/1538-4365/227/2/21},
  \href {https://ui.adsabs.harvard.edu/abs/2016ApJS..227...21T} {227, 21}

\bibitem[\protect\citeauthoryear{{Trac}, {Bode}  \& {Ostriker}}{{Trac}
  et~al.}{2011}]{trac11}
{Trac} H.,  {Bode} P.,   {Ostriker} J.~P.,  2011, \mn@doi [\apj]
  {10.1088/0004-637X/727/2/94}, \href
  {https://ui.adsabs.harvard.edu/abs/2011ApJ...727...94T} {727, 94}

\bibitem[\protect\citeauthoryear{{Trac}, {Chen}, {Holst}, {Alvarez}  \&
  {Cen}}{{Trac} et~al.}{2022}]{trac22}
{Trac} H.,  {Chen} N.,  {Holst} I.,  {Alvarez} M.~A.,   {Cen} R.,  2022,
  \mn@doi [\apj] {10.3847/1538-4357/ac5116}, \href
  {https://ui.adsabs.harvard.edu/abs/2022ApJ...927..186T} {927, 186}

\makeatother
\end{thebibliography}



\appendix


\bsp	
\label{lastpage}
\end{document}